\documentclass{dune}
\pdfoutput=1

\input{common/preamble}

\begin{document}

\pagestyle{titlepage}


\pagestyle{titlepage}

\date{}

\title{\scshape\Large White Paper on New Opportunities at the Next-Generation Neutrino Experiments \\
\normalsize (Part 1: BSM Neutrino Physics and Dark Matter)
\vskip -10pt
}


\renewcommand\Authfont{\scshape\small}
\renewcommand\Affilfont{\itshape\footnotesize}

\author[1]{C.A.~Arg\"uelles}
\author[2]{A.J.~Aurisano}
\author[3]{B.~Batell}
\author[3]{J.~Berger}
\author[4]{M.~Bishai}
\author[5]{T.~Boschi}
\author[6]{N.~Byrnes}
\author[6]{A.~Chatterjee}
\author[6]{A.~Chodos}
\author[7]{T.~Coan}
\author[8]{Y.~Cui}
\author[9]{A.~de~Gouv\^ea$^{\thanks{Lead Editors}}$ }
\author[4]{P.B.~Denton}
\author[* 10]{A.~De~Roeck}
\author[11]{W.~Flanagan}
\author[12]{D.V.~Forero}
\author[13]{R.P.~Gandrajula}
\author[14]{A.~Hatzikoutelis}
\author[15]{M.~Hostert}
\author[6]{B.~Jones}
\author[16]{B.J.~Kayser}
\author[16]{K.J.~Kelly}
\author[17]{D.~Kim}
\author[10,18]{J.~Kopp}
\author[19]{A.~Kubik}
\author[20]{K.~Lang}
\author[21]{I.~Lepetic}
\author[16]{P.A.N.~Machado}
\author[22]{C.A.~Moura}
\author[6]{F.~Olness}
\author[23]{J.C.~Park}
\author[15]{S.~Pascoli}
\author[12]{S.~Prakash}
\author[6]{L.~Rogers}
\author[24]{I.~Safa}
\author[24]{A.~Schneider}
\author[25]{K.~Scholberg}
\author[26,27]{S.~Shin}
\author[28]{I.M.~Shoemaker}
\author[25]{G.~Sinev}
\author[6]{B.~Smithers}
\author[* 2]{A.~Sousa}
\author[29]{Y.~Sui}
\author[30]{V.~Takhistov}
\author[31]{J.~Thomas}
\author[2]{J.~Todd}
\author[16,32]{Y.-D.~Tsai}
\author[33]{Y.-T.~Tsai}
\author[* 6]{J.~Yu}
\author[4]{C.~Zhang}

\vspace{-0.5cm}
\affil[1]{Massachusetts Institute of Technology, Cambridge, MA, USA}
\affil[2]{University of Cincinnati, Cincinnati, OH, USA}
\affil[3]{University of Pittsburgh, Pittsburgh, PA, USA}
\affil[4]{Brookhaven National Laboratory, Upton, NY, USA}
\affil[5]{Queen Mary University of London, London, UK} 
\affil[6]{University of Texas, Arlington, TX, USA} 
\affil[7]{Southern Methodist University, Dallas, TX, USA}
\affil[8]{University of California, Riverside, CA, USA}
\affil[9]{Northwestern University, Evanston, IL, USA}
\affil[10]{CERN, Meyrin, Geneva, Switzerland}
\affil[11]{University of Dallas, Dallas, TX, USA}
\affil[12]{IFGW - UNICAMP, 13083-859, Campinas, SP, Brazil}
\affil[13]{Michigan State University, East Lansing, MI, USA}
\affil[14]{San Jose State University, San Jose, CA, USA}
\affil[15]{Durham University, Durham, UK}
\affil[16]{Fermi National Accelerator Laboratory, Batavia, IL, USA}
\affil[17]{University of Arizona, Tuscon, AZ, USA}
\affil[18]{Johannes Gutenberg-Universit\", Mainz, Germany}
\affil[19]{Texas A\&M University, College Station, TX, USA}
\affil[20]{University of Texas, Austin, TX, USA}
\affil[21]{Illinois Institute of Technology, Chicago, IL, USA}
\affil[22]{Federal University of ABC, Santo Andr\'e, SP, Brazil}
\affil[23]{Chungnam National University, Daejeon, Korea}
\affil[24]{University of Wisconsin, Madison, WI, USA}
\affil[25]{Duke University, Durham, NC, USA}
\affil[26]{Enrico Fermi Institute and University of Chicago, Chicago, IL, USA}
\affil[27]{IPAP and Yonsei University, Seoul, Korea}
\affil[28]{Virginia Tech University, Blacksburg, VA, USA}
\affil[29]{Washington University, St. Louis, MO, USA} 
\affil[30]{University of California, Los Angeles, CA, USA}
\affil[31]{University College, London, UK}
\affil[32]{Kavli Institute for Cosmological Physics, University of Chicago, Chicago, IL, USA}
\affil[33]{SLAC National Laboratory, Menlo Park, CA, USA}

\maketitle

\renewcommand{\familydefault}{\sfdefault}
\renewcommand{\thepage}{\roman{page}}
\setcounter{page}{0}

\pagestyle{plain} 
\clearpage
\textsf{\tableofcontents}





\renewcommand{\thepage}{\arabic{page}}
\setcounter{page}{1}

\pagestyle{fancy}

\fancyhead{}
\fancyhead[RO]{\textsf{\footnotesize \thepage}}
\fancyhead[LO]{\textsf{\footnotesize \rightmark}}

\fancyfoot{}
\fancyfoot[RO]{\textsf{\footnotesize White Paper}}
\fancyfoot[LO]{\textsf{\footnotesize New Opportunities at the Next-Generation Neutrino Experiments}}
\fancypagestyle{plain}{}

\renewcommand{\headrule}{\vspace{-4mm}\color[gray]{0.5}{\rule{\headwidth}{0.5pt}}}



\clearpage

\section*{Executive Summary}
\label{sec:exec-summary}

With the advent of a new generation of neutrino experiments which leverage high-intensity neutrino beams for precision measurements, it is timely to explore physics topics beyond the standard neutrino-related physics. 
Given that the realm of beyond the standard model (BSM) physics  has been mostly sought at high-energy regimes at colliders, such as the LHC at CERN, the exploration of BSM physics in neutrino experiments will enable complementary measurements at the energy regimes that balance that of the LHC. 
This, furthermore, is  in concert with new ideas for high-intensity beams for fixed target and beam-dump experiments world-wide, e.g., those at CERN.

The combination of the high intensity proton beam facilities and massive detectors for precision neutrino oscillation parameter measurements and for CP violation phase measurements will help make BSM physics reachable even in low energy regimes in the accelerator based experiments.
Large mass detectors with highly precise tracking and energy measurements, excellent timing resolution, and low energy thresholds will enable the searches for BSM phenomena from cosmogenic origin, as well.
Therefore, it is also conceivable that BSM topics in the next-generation neutrino experiments could be the dominant physics topics in the foreseeable future, as the precision of the neutrino oscillation parameter and CPV measurements continues to improve.

In this spirit, this white paper provides a review of the current landscape of BSM theory in neutrino experiments in two selected areas of the BSM topics - dark matter and neutrino related BSM - and summarizes the current results from existing neutrino experiments to set benchmarks for both theory and experiment.
This paper then provides a review of upcoming neutrino experiments throughout the next 10 -- 15 year time scale and their capabilities to set the foundation for potential reach in BSM physics in the two aforementioned themes.
One of the most important outcomes of this white paper is to ensure theoretical and simulation tools exist to carry out studies of these new areas of physics, from the first day of the experiments, such as DUNE and Hyper-K.
Tasks to accomplish this goal, and the time line for them to be completed and tested to become reliable tools in a timely fashion, are also laid out.

\cleardoublepage

\section*{Preamble}
\label{sec:exec-preamble}
The authors of this paper primarily consist of those who participated in the {\it New Opportunities at Next-Generation Neutrino Experiments} workshop held on the campus of the University of Texas at Arlington.
Given the anticipated and growing importance of BSM topics that can be explored in the next-generation neutrino experiments, the authors of this white paper are strong advocates of these topics and aim to become a {\it de facto} working group for the upcoming decadal study of the APS Division of Particles and Fields.
This white paper is the first step to accomplish these goals, and provides a benchmark for ourselves to check progress toward making BSM physics a primary, yet complementary, topical area to precision neutrino oscillation parameter measurements.
The authors plan to follow up this white paper and workshop with the subsequent series to ensure continued improvement and broadening of the BSM topics accessible to neutrino experiments, and to draw increasing attention to this field throughout the future.

\section{Introduction}
\label{sec:exec-introduction}

Beyond the Standard Model physics is often associated with the realm of high-energy colliders, such as the \tevatron and the \lhc.
However, there is also an increasing interest to 
search for new physics in the lower mass regime
with high-intensity beams for fixed-target and beam-dump experiments. Similar opportunities can be explored at the planned high-intensity 
neutrino beam facilities.

The current bread-and-butter physics topics in neutrino experiments which aim to perform precision neutrino oscillation parameter measurements and the CP violation phase measurements require control over both statistical and systematic uncertainties. Thus, the next-generation neutrino experiments, such as \dune~\cite{Abi:2018rgm,Abi:2018dnh,Abi:2018alz} and \hyperk~\cite{Abe:2018uyc}, equipped with high-intensity proton beams, large detector mass, and precision measurement capabilities make \bsm physics more reachable.
In order to explore these newly accessible physics topics and to prepare tools and infrastructure necessary for physics from day 1, the Workshop on {\it New Opportunities at the Next-Generation Neutrino Experiments} was held at the University of Texas at Arlington on April 12 and 13, 2019.

This white paper is a tangible outcome to document the work performed at the workshop.
This white paper also establishes the benchmark for 1) what \bsm results look like at present, 2) what results of \bsm physics are expected in the next-generation neutrino experiments on what time scale, and 3) what tasks need to be accomplished in order to enable \bsm physics on day 1 and on what time scale these tasks must be accomplished.
The structure of this white paper reflects the three benchmark issues listed above. 
Various working groups consisting of a combination of theorists and experimentalists are expected to form and accomplish the tasks in a timely manner and to keep track of progress.
These benchmarks will also help the community to review progress at the subsequent workshops for continued advancement toward physics on day 1.

Of the many accessible \bsm physics topics at neutrino experiments, this white paper is centered on two themed topics, as to facilitate focused discussion and limit the scope of documentation generated on them to a manageable level.
The two themed topics covered in this white paper are searches of dark matter and \bsm topics involving neutrinos.
It is anticipated that other \bsm topics not covered in this white paper will be addressed in subsequent workshops in the series and the associated white papers.
It is also expected that a variation of this white paper will become part of the upcoming \snowmass strategy document.

\section{Landscape of BSM Physics at Neutrino Experiments}
\label{sec:exec-landscape}


\subsection{Neutrino BSM}
Despite their abundance, there are still many unknowns surrounding neutrinos. Most egregious is the fact that the Standard Model cannot account for neutrino masses. This fact guarantees some level of BSM physics in the neutrino sector. However, given the difficulties in detecting many neutrino properties, a number of additional possibilities exist as well, including the coupling of neutrinos to new force carriers. Many, but not all, BSM possibilities can be introduced in the framework of effective field theory.  In this framework, the new BSM physics energy scale $\Lambda$ is assumed to be large compared to accessible energies, $E \ll \Lambda$. Then BSM physics can be encoded as non-renormalizable interactions among SM particles which are suppressed by a new physics scale $\Lambda$   
\begin{equation}
\mathcal{L}_{{\rm eff}} = \mathcal{L}_{{\rm SM}} + \frac{c^{d=5}}{\Lambda} \mathcal{O}^{d=5} + \frac{c^{d=6}}{\Lambda^{2}} \mathcal{O}^{d=6}+ \cdots,
\label{eq:eft}
\end{equation}
where the sum continues, but terms are increasingly suppressed by additional powers of $\Lambda$.
In Eq.~(\ref{eq:eft}) it is well-known that there is one unique dimension-5 interaction: the Weinberg operator $\mathcal{O}^{d=5} = (LH)^{2}$, where $L$ and $H$ are the lepton and Higgs doublets, respectively~\cite{Weinberg:1979sa}. After the Higgs acquires its vacuum expectation value (VEV), a Majorana mass for neutrinos is generated. Ultraviolet completions of the Weinberg operator and the so-called see-saw mechanism for generated in neutrino masses are categorized based on the type of heavy particles introduced to generate $\mathcal{O}^{d=5}$. These are the Type-I see-saw, with heavy neutral fermions~\cite{Minkowski:1977sc,Yanagida:1979as,GellMann:1980vs,Mohapatra:1979ia}; the Type-II see-saw, with a scalar $SU(2)_L$ triplet; and the Type-III see-saw with fermionic $SU(2)_L$ triplets. Further extensions, e.g. inverse, extended and linear see-saw~\cite{GonzalezGarcia:1990fb,Pilaftsis:1991ug}, or new interactions, invoke a more extended fermionic or gauge sector leading to a richer phenomenology. 

\begin{itemize}
\item {\bf Sterile Neutrinos--}
The addition of neutral fermions, singlets with respect to the Standard Model gauge interactions, constitutes a minimal extension of the SM. As shown above, these can provide a mechanism for generating the light neutrino masses observed in experiments. However, there is no strong theoretical guidance on what mass such sterile neutrinos should have -- some studies propose very light (sub-eV) sterile neutrinos, whereas others propose very heavy sterile neutrinos with masses on the scale of grand unification, inspired by $SO(10)$-based models. In the latter case, decays of heavy neutrinos may be responsible for the baryon assymetry of the universe via leptogenesis~\cite{Fukugita:1986hr,Covi:1996wh,Pilaftsis:1997jf,Buchmuller:1997yu}. Between extremely high scales and low scales, many studies have searched for MeV-GeV sterile neutrinos via meson decays and in collider environments. While a qualitatively different mechanism of leptogenesis compared to the high-scale scenario, sterile neutrinos with these masses may also provide a solution to the baryon asymmetry of the universe~\cite{Akhmedov:1998qx,Asaka:2005pn}. Hints for low-scale sterile neutrinos have been present for over two decades, providing a potential solution for anomalous short-baseline neutrino oscillation results. These hints are in tension with searches for other channels of neutrino oscillations -- see, for instance, Refs.~\cite{Dentler:2018sju,Diaz:2019fwt,Boser:2019rta}, for a comprehensive review of this situation.
If these sterile neutrinos have additional interactions, e.g., with new dark gauge bosons~\cite{Okada:2014nsa,Ballett:2019pyw}, their phenomenology could be significantly different and new type of signatures could be present~\cite{Bertuzzo:2018itn,Ballett:2018ynz}.
\item {\bf Non-Standard Neutrino Interactions (NSIs)--}
Neutrino NSI (see, e.g., Ref.~\cite{Farzan:2017xzy}) refers to dimension-six four-fermion operators between neutrinos and matter that alter the production, detection, or propagation of neutrinos. These appear in two broad classes: charged current (CC)  $\mathcal{L} \supset (\ell_{\alpha} \gamma_{\mu} \nu_{\beta})(f\gamma^{\mu} f')$ and neutral current (NC) $\mathcal{L} \supset (\nu_{\alpha} \gamma_{\mu} \nu_{\beta})(f\gamma^{\mu} f)$, where $\ell_{\alpha}$ is a charged lepton and $f,f'$ are any SM fermions. CC NSI lead to modifications in the production and detection rates of neutrinos, including flavor-violating processes where, for instance, a $\nu_e$ strikes a target and produces a $\tau^-$ charged lepton. NC NSI, when the SM fermions in the new four-fermion operator are up- or down-quarks or electrons, modify the matter potential that neutrinos experience during propagation. The four-fermion NSI operators are dimension-six, and their couplings are typically expressed relative to $G_F$, the Fermi constant. Neutral Current NSI are less constrained at present than charged current ones, however DUNE will be sensitive to couplings at level of $\mathcal{O}(0.05-0.5)$ because of the large amount of matter  between neutrino production and detection~\cite{deGouvea:2015ndi,Coloma:2015kiu}. While the structure of the four-fermion operators is not UV complete, detectably large NSI arise in many light $Z'$ models~\cite{Farzan:2015doa,Farzan:2015hkd,Babu:2017olk,Denton:2018xmq}.  Light $Z'$ models include other notable BSM phenomenology such as trident events~\cite{Altmannshofer:2014pba,Altmannshofer:2019zhy,Ballett:2019xoj} and modified production channels for heavy sterile neutrinos~\cite{Batell:2016zod}. In addition to DUNE, future bounds on NC NSI will come from atmospheric neutrino observations at IceCube~\cite{Aartsen:2017xtt} and Hyper-Kamiokande~\cite{Kelly:2017kch} and at the COHERENT experiment~\cite{Farzan:2017xzy,Akimov:2017ade,Coloma:2017ncl}. 

\item {\bf Neutrino Interactions with Dark Matter--} Given that both neutrinos and dark matter (DM) require BSM physics, it is natural to consider models in which they are connected. Models with light DM and light mediators have been explored, with detectable modifications to neutrino oscillation probabilities~\cite{Berlin:2016woy,Krnjaic:2017zlz,Brdar:2017kbt,Capozzi:2017auw,Liao:2018byh,Capozzi:2018bps,Ge:2019tdi}.

\item {\bf Neutrino Dipole Operators--} Although neutral under the electromagnetic force, neutrinos may couple to the photon via a higher dimensional operator such as a dipole moment. Bounds from neutrino-electron scattering data provide the strongest direct laboratory bounds, and depend strongly on the neutrino flavor used. For example, the GEMMA experiment observation of $\bar{\nu}_{e}-e$ scattering places the strongest bounds on an electron flavor dipole moment~\cite{Beda:2012zz}, while LSND~\cite{Auerbach:2001wg} and DONUT~\cite{Schwienhorst:2001sj} provide the best bounds on $\mu$-flavor and $\tau$-flavor dipoles, respectively. Currently Borexino and Super-K use the shape of the electron spectrum in solar neutrino data to provide bounds on an effective magnetic moment~\cite{Borexino:2017fbd,Liu:2004ny}. Additionally, magnetic moments may provide a portal to sterile neutrinos via active-to-sterile transition moments. Existing bounds have been studied in Refs.~\cite{Magill:2018jla,Coloma:2017ppo}. Future bounds on these transition moments may come from IceCube~\cite{Coloma:2017ppo}, SHiP, SBND~\cite{Magill:2018jla}, and DM direct detection experiments~\cite{Shoemaker:2018vii}.  
\end{itemize}

\subsection{Dark Matter}\label{sec:DM}
In recent years, the search for DM with neutrino facilities has become a blooming research area presenting many new opportunities. 
Next-generation neutrino experiments offer particular advantages over existing experiments, including sensitivity to energetic DM scattering, large volumes, and high-intensity beams for potential DM production. We briefly summarize and classify these opportunities as follows, with further details in Sec.~\ref{sec:exec-prospects}.

 \textbf{I. Indirect Detection from DM annihilation/decay.}
    This is the traditional indirect detection of DM using neutrino experiments. The DM signal is revealed as a flux of neutrinos from DM annihilation/decay in DM-rich astrophysical objects such as the Galactic Center (GC) or the Sun. \superk and \icecube have existing/ongoing searches in this direction, as explained in Sec.~\ref{sec:exec-current}.
 
 \textbf{II. Direct Detection of cosmogenic DM.}
    Conventional, direct detection searches for GeV-scale DM assume that the relic DM is moving slowly in the galaxy, $v \approx 10^{-3} c$. When a DM particle interacts with the detector, its energy deposition is small. The direct detection experiments are therefore aimed to be low-threshold, low-background environments. In contrast, because of their large volume, neutrino detectors tend to have much higher thresholds, and therefore are insensitive to the scattering of relic DM off detector targets. If mechanisms exist beyond this paradigm, then neutrino detectors are advantageous thanks to their large volume.
 
    \begin{itemize}
    \item \textbf{Boosted DM and related variations}.
    Boosted DM (BDM) is a novel solution to the above situation, proposing that a small, non-thermal component of DM is present in nature today and is relativistic~\cite{Agashe:2014yua, Huang:2013xfa, Berger:2014sqa, Kong:2014mia, Kim:2016zjx}. This small component may be associated with the freeze out of the larger component of DM in a two-component model~\cite{Agashe:2014yua,Belanger:2011ww}, or associated in semi-annihilation scenarios with dark-sector symmetry like $\mathbb{Z}_3$~\cite{DEramo:2010keq}. Signatures of BDM vary from DM-induced nucleon decays~\cite{Huang:2013xfa}, decaying massive particles~\cite{Bhattacharya:2014yha, Kopp:2015bfa, Cui:2017ytb}, or energetic cosmic-ray-induced BDM~\cite{Yin:2018yjn, Bringmann:2018cvk, Ema:2018bih,Dent:2019krz}.
    
    The phenomenology of boosted DM has two features: a small flux and energetic final state electrons or hadrons upon BDM scattering on SM targets. Because of these, large volume neutrino experiments are generally most suitable for BDM detection~\cite{Agashe:2014yua,Berger:2014sqa,Kim:2016zjx, Kong:2014mia,Necib:2016aez,Alhazmi:2016qcs}, while complementary searches in WIMP direct detection experiments~\cite{Cui:2017ytb, Cherry:2015oca,  Giudice:2017zke} and the intermediate-volume surface-based neutrino experiments~\cite{Chatterjee:2018mej, Kim:2018veo} can possibly contribute, covering wide parameter regions.

    BDM signals may be similar to neutrino-related backgrounds, but considerations of the incoming direction of the DM may assist in reducing these backgrounds~\cite{Agashe:2014yua,Berger:2014sqa}. Additionally, some BDM models feature distinct signatures after the BDM scatters, such as scattering into an unstable state which decays promptly and semi-visibly into SM particles~\cite{Kim:2016zjx, Giudice:2017zke, Chatterjee:2018mej,Kim:2019had}.
        \item \textbf{Self-destructing DM.}
        Another example of cosmogenic DM able to be probed in neutrino experiments is self-destructing DM~\cite{Grossman:2017qzw,Eby:2019mgs}. 
        A dark-matter component can have a transition from a long-lived state to a short-lived one by scattering off of material in the Earth. The latter then can decay to SM particles, e.g., back-to-back $e^+ e^-$, inside a large detector with a visible energy of order of the DM mass.
    \end{itemize}
\textbf{III. Beam-produced DM at accelerator neutrino facilities.}
    If new, weakly-coupled MeV-GeV particles exist, they can potentially be produced in the large number of proton-proton collisions at accelerator neutrino facilities~\cite{Batell:2009di}. Such DM is typically produced via decays of mesons, and can travel to the neutrino facility's near detector(s) and interact~~ \cite{deNiverville:2011it,deNiverville:2012ij,Dharmapalan:2012xp,Batell:2014yra,Dobrescu:2014ita,Soper:2014ska,Kahn:2014sra,deNiverville:2015mwa,Kuflik:2015isi,Coloma:2015pih,deNiverville:2016rqh,Izaguirre:2017bqb,Frugiuele:2017zvx,Kuflik:2017iqs,Jordan:2018gcd,deNiverville:2018dbu,Batell:2018fqo,deGouvea:2018cfv,DeRomeri:2019kic,Dutta:2019nbn}.
    
    A subset of this class of models is millicharged particles, in which the new particle interacts weakly with either the SM photon or a massless dark photon. Existing constraints on $\sim$MeV-GeV millicharged particles are driven by these types of searches at MiniBooNE and LSND~\cite{Magill:2018tbb}.




\section{BSM Physics Results from Current Neutrino Experiments}
\label{sec:exec-current}

Two distinct classes of BSM physics are considered: effects of BSM physics on neutrino oscillations and dark matter signatures in neutrino experiments.

\subsection{Effects of BSM Physics on Neutrino Oscillations}

\begin{figure}[t]
    \centering
    \includegraphics[height=3.5in,width=0.43\textwidth]{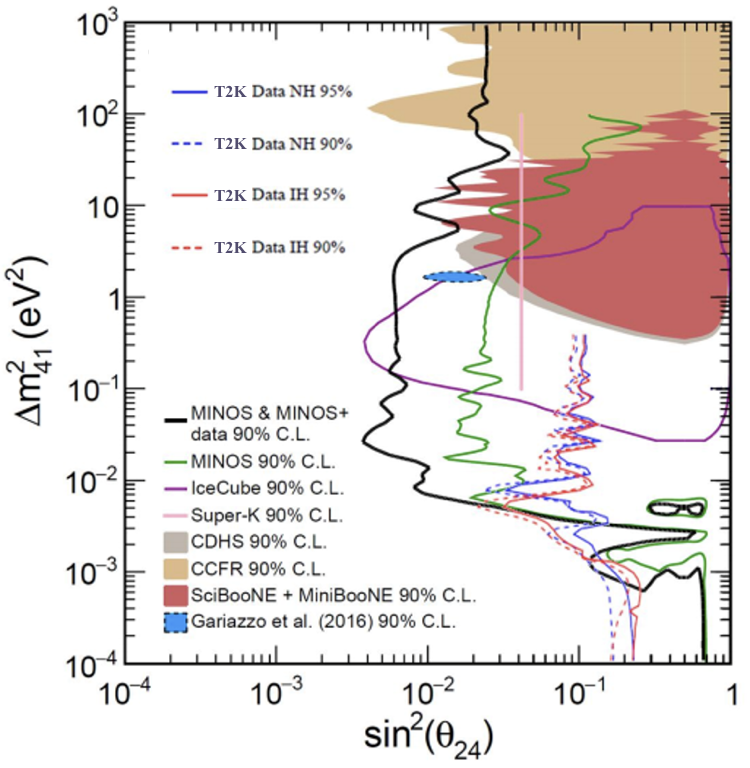}
    \includegraphics[height=3.7in,width=0.47\textwidth]{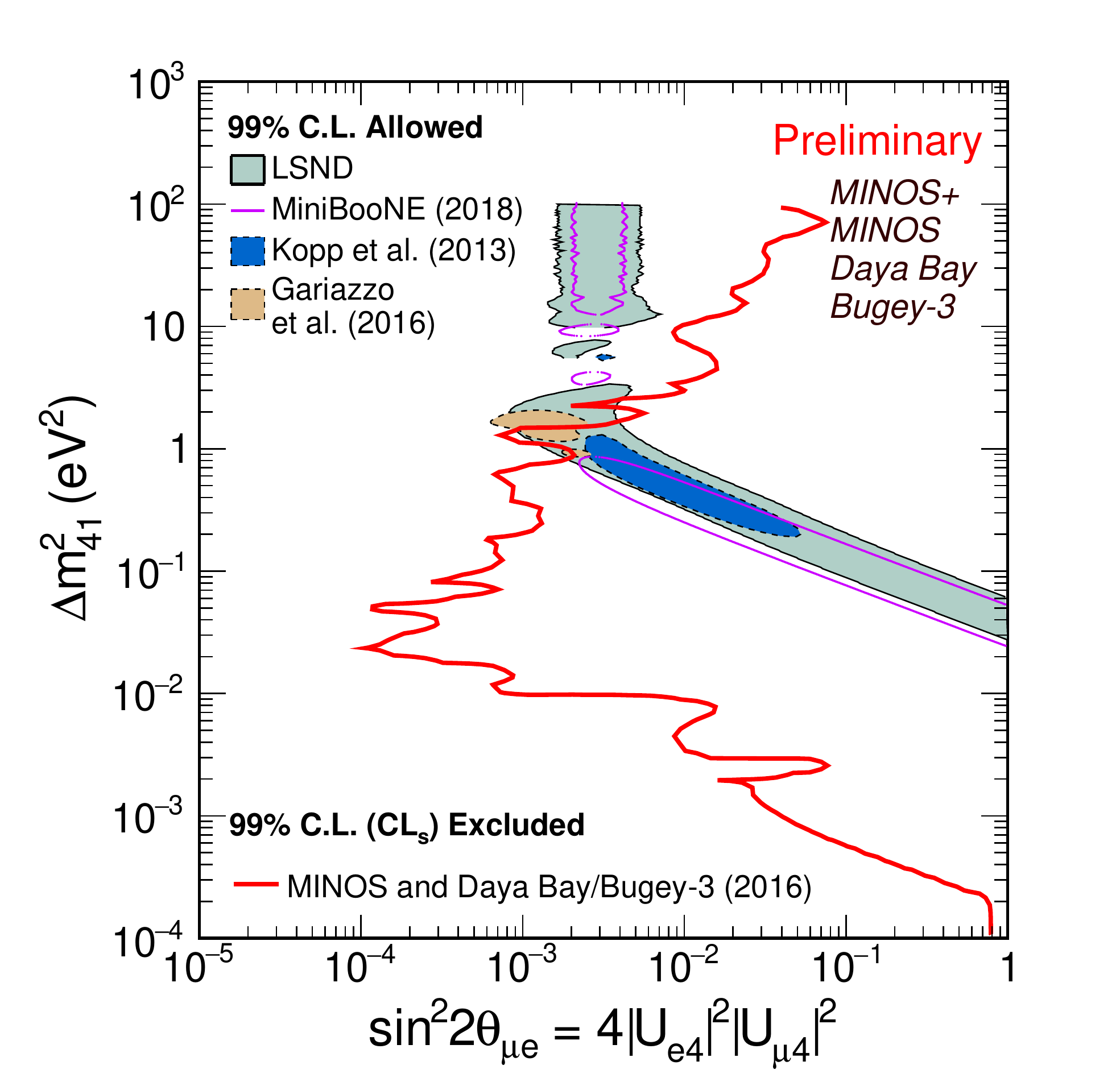}
    \caption{Left panel: Comparison of present exclusion limits from various experiments obtained through searches for disappearance of muon neutrinos into sterile species assuming a 3+1 model.  The Gariazzo et al. region represents a global fit to neutrino oscillation data~\cite{Gariazzo_2015}. 
    Right panel: The combined results of the disappearance measurements from \minosplus, \dayabay, and \bugey, compared to the appearance measurements from \lsnd and \miniboone.}
    \label{fig:disappearance}
\end{figure}

There are several categories whose signals would manifest themselves as deviations from the known measured ``standard model'' of neutrinos~\cite{Asaka:2005an,ASAKA200517}. It is clear that any search for deviations of this model is limited by our knowledge of the standard parameters and will always be so. 
The present lead in the search for sterile neutrinos, those which couple to standard neutrinos but not to the weak interaction, comes from disappearance experiments such as muon-neutrino accelerators and reactor anti-neutrino experiments, where unitarity is a necessary assumption. All the most precise measurements of the standard oscillation parameters have been made by disappearance experiments as shown in the left panel of Figure~\ref{fig:disappearance}. The \lsnd and \miniboone anomalies are expected to be elucidated by \microboone due to its unprecedented event reconstruction capabilities. After the recent measurement from \minosplus and \icecube are combined with unitarity constraints (see e.g.~\cite{Parke:2015goa}), most of the favored parameter space to explain \lsnd and \miniboone, with a sterile neutrino, is now disfavored as shown in the right panel of Figure~\ref{fig:disappearance}. Addressing the apparent excess of electron events appearing in the muon-neutrino beam at \miniboone and \lsnd is also the main goal for the future SBN program at Fermilab.

Searches for Non-Standard Interactions (NSI) have also been carried out using different neutrino event samples.
Different global analyses of the neutrino oscillation and neutrino scattering data have been used to search for and constrain the strength of NSI's ~\cite{Denton:2018xmq,Coloma:2017egw,Esteban:2018ppq}.
More details are presented and discussed in Sec.~\ref{sec:nsi}.
  
\begin{figure}
    \centering
    \includegraphics[width=0.7\textwidth]{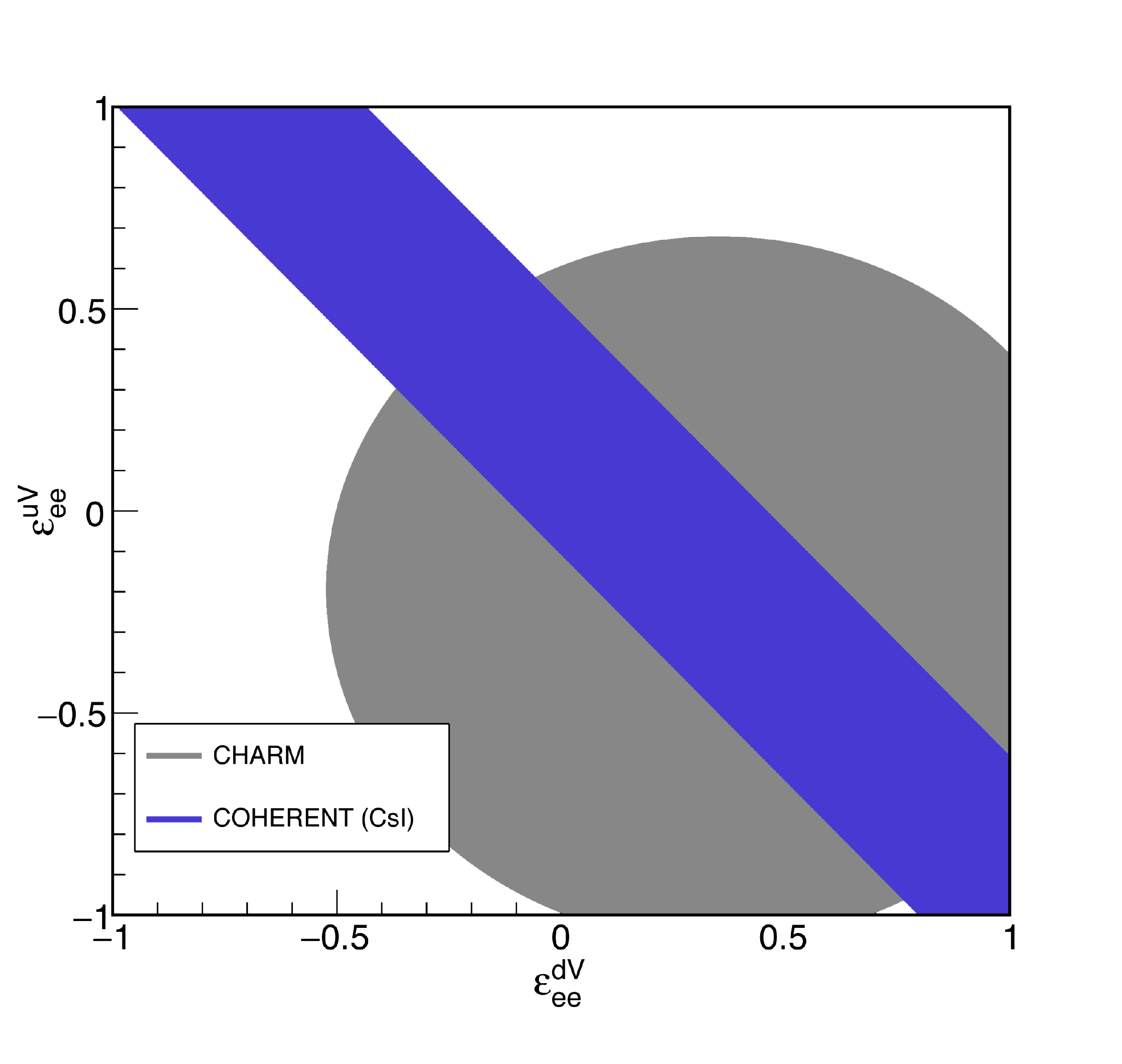}
    \caption{The \coherent NSI constraint compared to the previous allowed region from CHARM from \cite{Akimov:2017ade}.}
    \label{fig:coherent_nsi}
\end{figure}

The first observation of coherent elastic neutrino-nucleus scattering (CE$\nu$NS)~\cite{Akimov:2017ade} was performed by the \coherent collaboration~\cite{Akimov:2018ghi} in 2017 using $\pi$-decay-at-rest neutrinos produced by the Spallation Neutron Source (SNS) at ORNL and a 14.6-kg CsI detector located 19.3 m from the SNS target. Using the same data, \coherent calculated limits in the $\epsilon^{dV}_{ee}$ and $\epsilon^{uV}_{ee}$ NSI parameter space (setting the rest of the NSI couplings to 0) at 90\% confidence level (the blue region in figure \ref{fig:coherent_nsi}) that significantly improved on the previous limit (the CHARM constraint is the gray region in Figure~\ref{fig:coherent_nsi}) and is working on constraining other NSI couplings. NSI studies (as well as studies of other BSM topics) with the published COHERENT data have also been performed independently~\cite{Coloma:2017ncl,Dent:2017mpr,Liao:2017uzy,Abdullah:2018ykz,Denton:2018xmq,Dutta:2019eml}, including exotic NSI inducing DM$\leftrightarrow \nu$ conversion via scattering~\cite{Brdar:2018qqj,Dror:2019onn}.


\subsection{Dark Matter Signatures in Neutrino Experiments}

The production of DM in the scope of this white paper can be categorized as cosmogenic and accelerator-produced DM, as introduced in Sec.~\ref{sec:DM}.
The cosmogenic DM signatures can be further split into two subsets: the search for neutrinos produced by WIMP-WIMP annihilation and the search for ``Boosted'' (relativistic) Dark Matter which would interact directly in the neutrino detectors. \superk~\cite{Kachulis:2017nci} and \icecube~\cite{PhysRevLett.119.201801} lead these searches presently, while the first search for inelastic BDM (iBDM) models, as described in~\ref{sec:bdm}, has been performed in \cosine~\cite{Ha:2018obm}. \kamland has the potential to contribute, but presently has no results. 

\begin{figure}[t]
    \centering
    \includegraphics[height=6.5cm,width=0.46\textwidth]{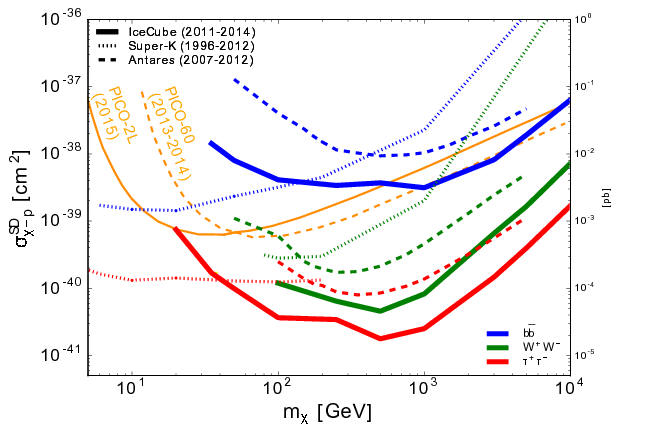}
    \includegraphics[height=6.5cm,width=0.4\textwidth]{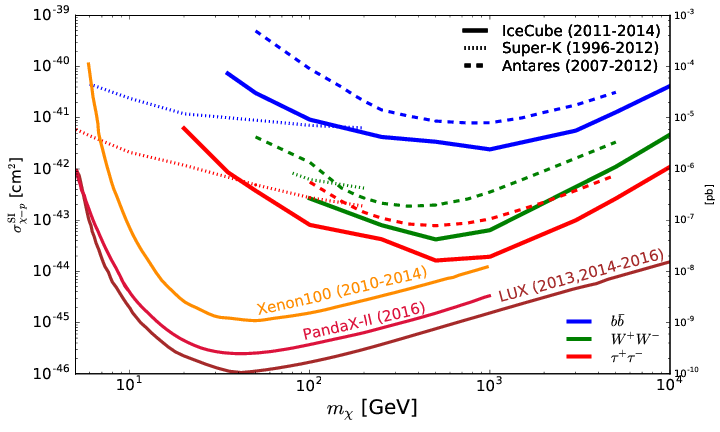}
    \caption{Current limit on high-energy neutrinos produced by spin-dependent proton scattering (left panel) and spin-independent nucleon scattering (right panel) of dark matter in the Sun~\cite{Scott:2017dki}.}
    \label{fig:spin_dependent}
\end{figure}

Comprehensive comparisons of the WIMP-WIMP annihilation searches to direct WIMP searches can be found in the left and right panels of Figure~\ref{fig:spin_dependent}
for spin-dependent and spin-independent WIMP scenarios, respectively. These searches are possible because DM can accumulate in the Sun to such high densities that DM annihilation can occur. Depending on the models, different spectra of high-energy neutrinos are produced, which can be searched for in terrestrial neutrino detectors. See Sec.~\ref{sec:dmnu} for further details of these searches.

\begin{figure}[th!]
    \centering
    \includegraphics[height=12cm]{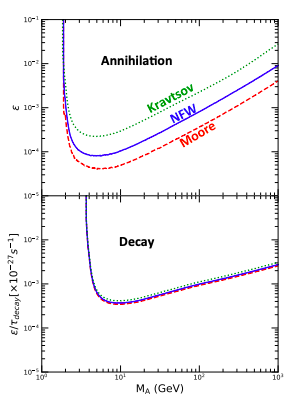}
    \includegraphics[height=11.5cm]{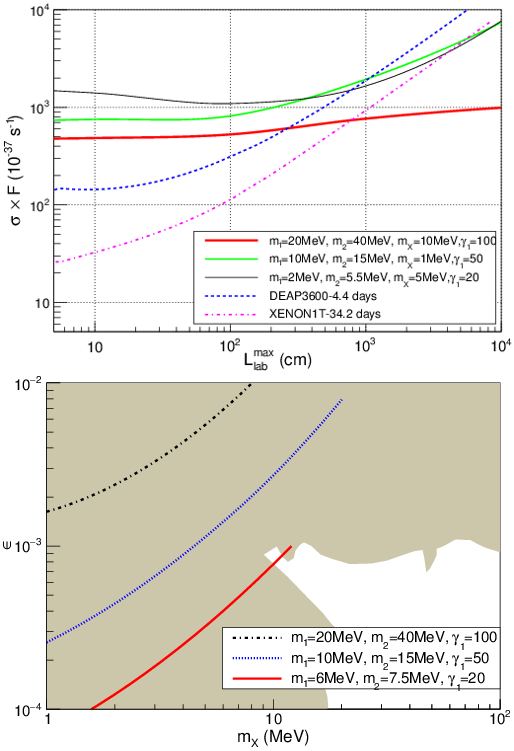}
    \caption{Left panel: 90\% Confidence Interval upper limits from the \superk result, with the mass of the BDM, $m_B$ = 200~MeV, the mass of the dark photon, $m_{\gamma}'$ = 20~MeV, and the coupling between the BDM and the dark photon, $g'$=0.5, for BDM produced by annihilation (top) and decay (bottom)~\cite{Kachulis:2017nci}.
    Right panel: Measured 90\% Confidence Interval upper limits (lines) from 59.5 days of \cosine data in the L$_{\rm lab}^{\max}$-$\sigma$ plane (upper panel) in terms of the mass of a mediator, $m_\chi$, and the mixing parameter between the standard model particle and the dark sector particle, $\epsilon$, assuming the mediator to be a dark photon (lower panel) are presented for three different benchmark models. In the upper panel the results are compared with the expected experimental sensitivities of XENON1T and DEAP-3600 calculated in Ref.~\cite{Giudice:2017zke}, while in the lower panel
    done with the currently excluded parameter space (shaded region) from direct dark photon search experiments E141~\cite{PhysRevLett.59.755}, NA48~\cite{Batley:2015lha}, NA64~\cite{PhysRevLett.120.231802}, \babar~\cite{Lees:2014xha}, and bounds from the electron anomalous magnetic moment $(g-2)_e$~\cite{Davoudiasl:2014kua}.}
    \label{fig:bdm}
\end{figure}

As shown in the left panel of Figure~\ref{fig:bdm}~\cite{Kachulis:2017nci},
the BDM limit is currently held by \superk, where the coupling strength of BDM to electrons is probed
with the data corresponding to 161.9 kiloton-years exposure.
The BDM is assumed to originate in the Galactic Center or the Sun, and the energy of the scattered electrons ranges from 100~MeV to 1~TeV.
The right panel of Figure~\ref{fig:bdm}~\cite{Ha:2018obm}, on the other hand, shows the recent result of the iBDM search from \cosine.

Searches for low-mass DM produced in accelerator beams have been demonstrated by \miniboone~\cite{Aguilar-Arevalo:2018wea}, and an example of the parameter space probing can be found in Figure~\ref{figure:MB-plot}.
A dedicated beam-dump configuration, where the proton beam is steered past the beryllium target, was deployed to greatly eliminate the neutrino flux, which is the main background for this search.


\FloatBarrier
\section{Next-Generation Neutrino Experiments and Their Capabilities}
\label{sec:exec-nextgen}

\subsection{Neutrino Sources and BSM Physics}

\begin{figure}[!htbp]
    \centering
    \includegraphics[width=0.35\linewidth]{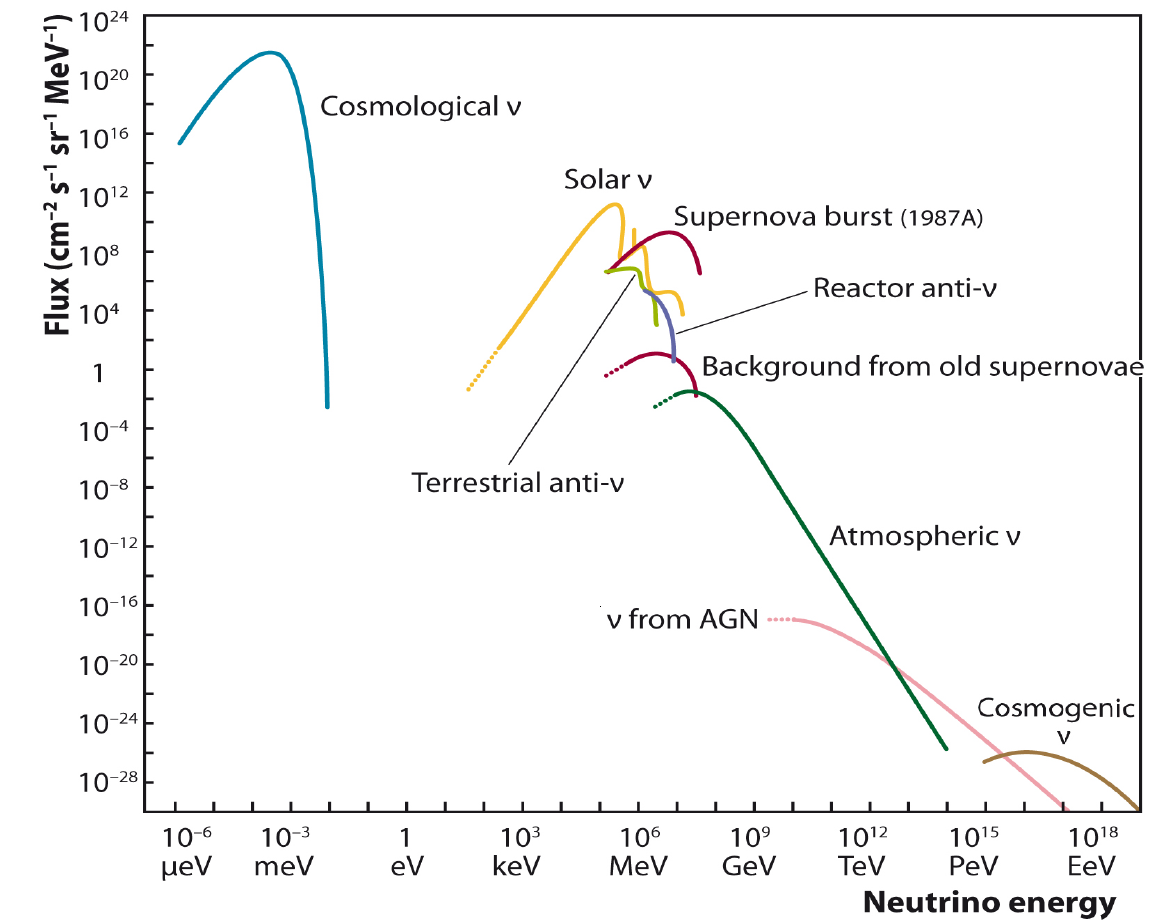}
    \includegraphics[width=0.4\linewidth]{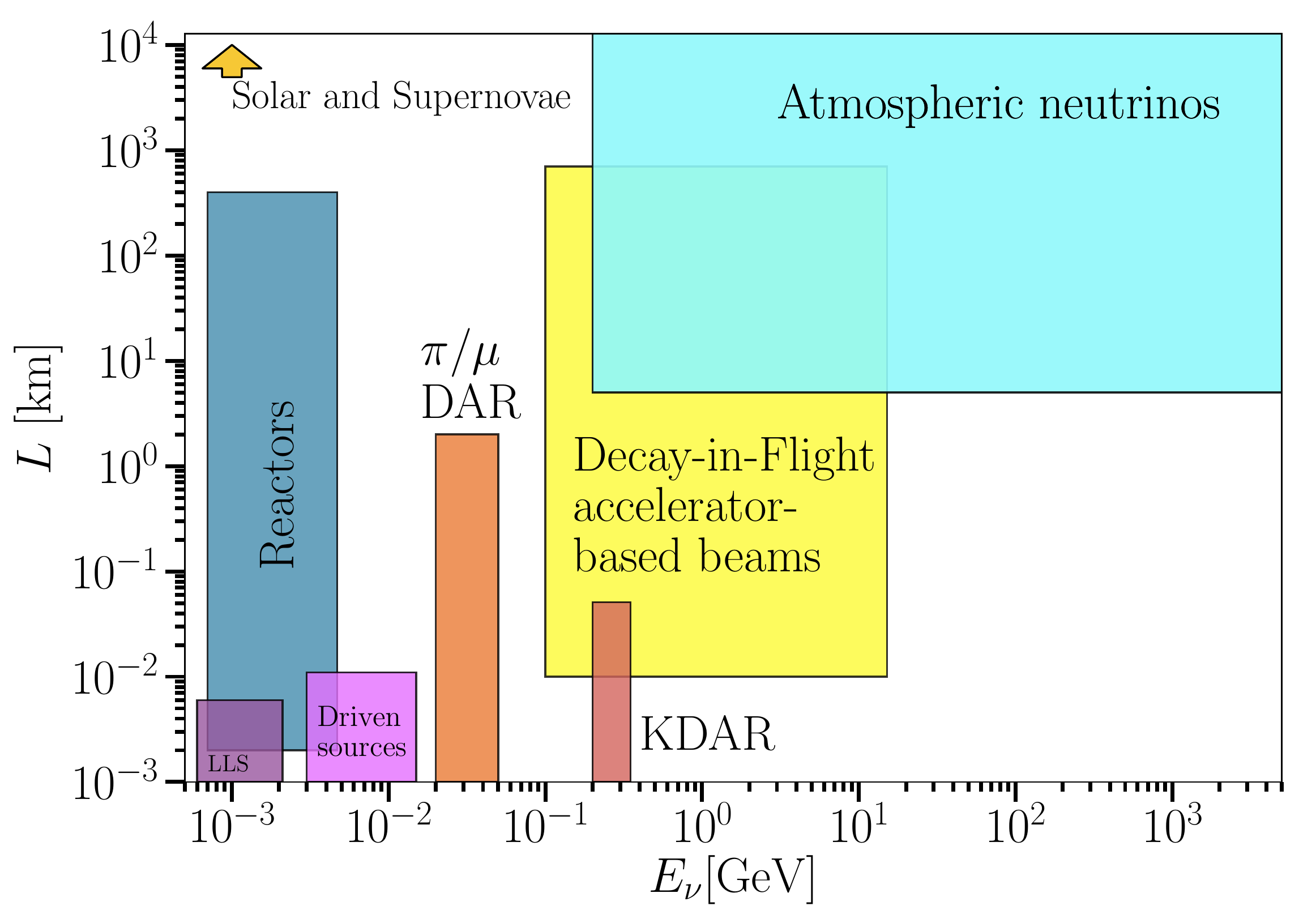}
    \caption{Left: Neutrino fluxes from different sources as a function of the neutrino energy~\cite{Katz:2011ke}. Right: Neutrino sources used in on-going and next generation neutrino experiments with the sources have been arranged as a function of the neutrino energy and typical distance from neutrino production to detector~\cite{Diaz:2019fwt}.}
    \label{fig:neutrino_sources}
\end{figure}

New physics models accessible to next-generation neutrino experiments are specific to the neutrino source used. Neutrino sources are either man-made (accelerators, nuclear reactors, or radioactive sources) or natural (geological, solar, atmospheric, supernova, cosmogenic, or big-bang); see Fig.~\ref{fig:neutrino_sources}. A short survey of neutrino sources and their characteristics are discussed in this section.  

\subsubsection{Accelerator Sources}

Accelerator sources of neutrinos can be divided into four categories: 
\begin{enumerate}
    \item {\bf Pion decay-in-flight:} These sources are primarily from high-intensity GeV-scale proton accelerators which produce beams of high-purity $\nu_\mu$ or $\bar{\nu}_\mu$ in the hundreds of MeV to hundreds of GeV energy range. Examples of current and future neutrino sources are the 8 GeV Booster Neutrino Beam (BNB) at Fermilab ($\sim$ 800 MeV $\nu_\mu$)~\cite{bnb1,bnb2}; the 60-120 GeV highly-tunable and configurable Neutrinos from Main Injector (NuMI) beam~\cite{Adamson:2015dkw}; the 30 GeV J-PARC neutrino beamline (600 MeV $\nu_\mu$ off-axis)~\cite{Abe:2011ks}; the proposed LBNF beamlines at Fermilab (wide-band on-axis $\nu_\mu$ beams or narrow off-axis in the range of a few to tens of GeV)~\cite{Marciano:2001tz,Strait:2016mof}; and the proposed neutrino beam from the Protvino accelerator complex~\cite{Akindinov:2019flp} which uses the U-70 accelerator which can produce a 90 kW proton beam with energies up to 70 GeV resulting in a 5 GeV neutrino beam at an 204 mrad off-axis detector. BSM physics searches from these sources are focused on searches for new effects in $\nu_\mu$ oscillations beyond the three-flavor paradigm at both short and long baselines. These sources are also powerful tools to search for dark matter or heavy neutrinos produced in the proton beam target or decay pipe.
    \item {\bf Pion decay-at-rest:} Pion decay-at-rest in beam dumps are sources of muon-type neutrinos in the range of a few tens of MeV. As the power of proton accelerators grows, the intensity of neutrinos from these sources has increased markedly, opening up new opportunities for new physics searches in $\nu_\mu$ oscillation searches (for e.g. OscSNS ~\cite{Garvey:2005pn} or DAE$\delta$ALUS~\cite{Alonso:2010fs}), neutrino scattering ({\it{} e.g.} coherent $\nu$-nucleus scattering), and dark matter or other new particles produced in the beam dumps and detected by the neutrino detectors deployed nearby. Studies and realizations of muon decay-at-rest and kaon decay-at-rest beams are also under serious investigation (see, for example, JSNS$^2$~\cite{Ajimura:2017fld}).
    \item {\bf Isotope decay-at-rest:} Isotope decay-at-rest experiments use the beta-decay of unstable isotopes, produced by a powerful accelerator, to generate a high-intensity and high-purity $\bar \nu_e$ beam. Current proposals include IsoDAR with $^8$Li which is in the R\&D phase~\cite{Alonso:2017fci}, as well as BEST with $^{51}$Cr~\cite{Barinov:2019vmp}, BEST-2 with $^{65}$Zn \cite{Gavrin:2019rtr}, and SOX with $^{51}$Cr at Borexino~\cite{Borexino:2013xxa}.
    \item {\bf Muon decay-in-flight:} Muons produced from pion decay-in-flight at high-power proton accelerators can be captured and accelerated in a muon storage ring. The decay of $\mu \rightarrow e \bar{\nu}_e \nu_\mu$ produces a source of equal parts $\nu_e$ and $\nu_\mu$ with a precisely known spectrum for both. Short-baseline experiments, such as the proposed nuSTORM~\cite{Adey:2013pio}, can use these beams to search for sterile neutrinos. nuSTORM does not require muon cooling and is therefore in principal technically feasible in the near future. With muon cooling - which is still in the R\&D phase - a {\it{} Neutrino Factory} can be constructed. This source will be a high-intensity muon decay-in-flight source 
    Unfortunately, there are many technical hurdles, as well as significant cost, to the realize Neutrino Factories appropriate use in long-baseline experiments and are therefore considered a far-future option.
\end{enumerate}

Other sources of accelerator neutrinos of note are the 400 GeV CERN Super Proton Syncroton (SPS) beam that was used to produce the $\sim 20$ GeV $\nu_\mu$ beam from the CNGS beamline~\cite{Acquistapace:1998rv}. This beam was used in experiments such as OPERA~\cite{Agafonova:2019npf} and ICARUS~\cite{Amerio:2004ze} experiments located in the Gran Sasso undeground laboratory. The SHiP experiment is proposing to use the SPS beam to generate a beam of $\nu_\tau$ from charm mesons decays~\cite{SHiP:2018yqc} and FASER~\cite{Feng:2017uoz} and FASERnu~\cite{Abreu:2019yak} will measure all three flavors produced at the ATLAS interaction point on the LHC.
This would allow a more accurate measurement of the $\nu_\tau$ and $\bar{\nu}_\tau$ cross-sections and enable searches for new physics in $\nu_\mu \rightarrow \nu_\tau$, since for these high-energy neutrino beams the charged-current tau neutrino cross section is not suppressed. 
SHiP also proposes to search for heavy neutrinos and dark matter produced in a specially designed dump by the SPS beam.
In addition, several other experiments have proposed measuring long-lived particles including neutrinos at the LHC such as FASER~\cite{Ariga:2019ufm,Abreu:2019yak}, which was recently approved and funded, and MATHUSLA~\cite{Curtin:2018mvb}.

\subsubsection{Reactor Sources}

\begin{figure}[!htbp]
    \centering
    \includegraphics{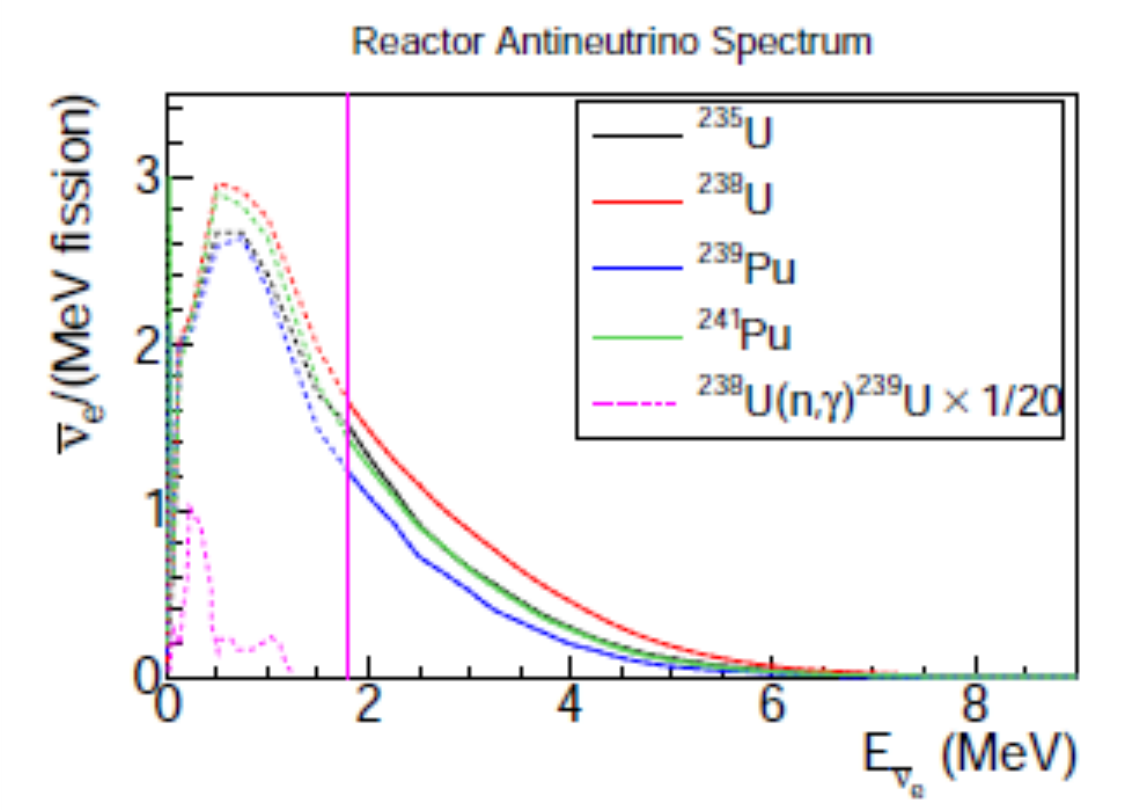}
    \caption{The $\bar{\nu}_e$ flux produced from the fission chain of the most common nuclear reactor isotopes~\cite{Qian:2018wid}. The inverse beta decay threshold is indicated by the magenta vertical line.}
    \label{fig:reactorflux}
\end{figure}

Nuclear reactors produce a flux of pure $\bar{\nu}_e$ with energy ranges from keV to a few MeV as shown in Fig.~\ref{fig:reactorflux}. Commercial reactors produce a large flux of neutrinos ($\approx 10^{20}$ neutrinos per second per gigawatt thermal power). While smaller research reactors are typically at the megawatt scale, though they benefit from a compact core.

Neutrinos with energies above 1.8 MeV can be detected through inverse beta decay (IBD), which has a well-known cross-section and enables high-precision measurement of the neutrino energy through its unique signature of a prompt positron annihilation followed by a delayed neutron capture. There are two main methods to predict the reactor anti-neutrino flux above the IBD treshold:
\begin{enumerate}
    \item {\it Ab-initio method:} The four main fission isotopes contributing to the reactor anti-neutrino flux in a commercial reactor are $^{235}$U, $^{239}$Pu, $^{238}$U, and $^{241}$Pu. The ab-initio method utilizes nuclear data tables and theoretical calculations to estimate the neutrino spectrum from beta-decays of all the fission daughters of each of the main four isotopes in a reactor~\cite{Hayes:2016qnu}. Over 6000 decay branches are included in the most up-to-date calculations. The precision of the predicted spectrum is limited by systematic uncertainties in the nuclear databases used and the typical precision of this method is around 10\%.
    \item {\it Conversion method:} The conversion method uses a fit to the measured total electron spectrum from the fission and subsequent beta decays of each of the four main isotopes. The theoretical beta decay spectrum of $\sim 30$ ``virtual'' branches with equidistant end point spacing and effective Z values is used to fit the measured electron spectrum~\cite{Huber:2011wv,Mueller:2011nm}. The neutrino spectrum can then be inferred from the fits. This method is limited by the experimental uncertainties in the beta decay spectrum measurements and theoretical uncertainties in the conversion method, in particular the spectrum shape associated with forbidden beta-decays. Typical precision of this method is $\sim 2-5\%$.
\end{enumerate}

Recent precision measurements of the reactor $\bar{\nu}_e$ spectrum, above the inverse-beta-decay threshold, by modern reactor experiments like Daya Bay, RENO, and Double Chooz\cite{Abe:2014bwa,Adey:2019ywk,RENO:2018pwo} have significantly improved our understanding of the reactor flux calculations and the reactor anti-neutrino anomaly (RAA)~\cite{Mention:2011rk}.
The new data indicate that incorrect modeling of the different fission isotope fluxes is favored as the main source of the RAA~\cite{Giunti:2019qlt}.
Very-short-baseline reactor experiments like PROSPECT~\cite{Ashenfelter:2018iov} and STEREO~\cite{Almazan:2018wln} will further improve the precision of the reactor anti-neutrino flux and spectrum measurements and increase the sensitivity to sterile searches with reactor neutrino oscillations.
Reactor neutrino sources are also hosts to many experiments measuring coherent neutrino-nucleus scattering (CEvNS) on a variety of nuclear targets.
Measurements of CEvNS, a process with a well-known cross-section, is a good place to look for BSM physics as demonstrated in~\cite{Coloma:2017ncl}. In addition to the experimental challenges of detecting low-energy neutrinos, the sensitivity to BSM at these experiments is limited by the reactor flux uncertainties.      

\subsubsection{Cosmogenic Neutrinos}

Cosmogenic neutrinos come from many different sources with a vast variation in characteristics and abundance. Some notable sources of cosmogenic neutrinos and their role in searches for BSM physics are summarized as follows.
\begin{enumerate}
\item{\bf Solar:} Solar neutrinos are produced by nuclear fusion reactions in the sun. Their propagation is unique in that the electron neutrinos experience a very-strong matter potential that modifies their oscillation via the Mikheyev-Smirnov resonance~\cite{Mikheev:1986gs}.
This, combined with their very-large flux at Earth, makes them an excellent tool to probe neutrino NSI and other exotic oscillation phenomena. The sun may also be a source of exotic particles, for example, exotic particle annihilation to boosted dark matter in the solar core, or of high-energy neutrinos through WIMP annihilation to neutrinos~\cite{Choi:2015ara,Aartsen:2016zhm,Adrian-Martinez:2016gti,Leane:2017vag,Nisa:2019mpb}. Solar neutrinos are also one of the limiting backgrounds to direct dark matter detection experiments and therefore need to be precisely understood as dark matter direct detection experiments approach the neutrino floor.
\item{\bf Atmospheric:} Atmospheric neutrinos are produced via the decays of charged pions, kaons, and charmed hadrons at the highest energies in the upper atmosphere of the Earth.  Sufficiently high energy ($\geq$ GeV) atmospheric neutrinos interacting in detectors provide boosted signatures with sufficient pointing information to infer the baseline from production, and topologies that allow for containment within the detector provide a capability to measure energy.  The ability to study dependencies upon both $L$ and $E$ in a continuous manner naturally makes atmospheric neutrinos a powerful tool to use to search for non-standard neutrino oscillation physics.  It is also possible to consider BSM particles co-produced with neutrinos within the air shower, which may themselves introduce signatures in experiments; see {\it e.g.}~\cite{Ahlers:2007js}.
\item{\bf High-energy astrophysical:} High-energy astrophysical neutrinos are produced in violent astrophysical environments such as active galactic nuclei, gamma ray bursts, blazars, and others~\cite{Ahlers:2018fkn}. They have both the longest baselines and highest energies of any detected neutrinos, which provides a unique handle on certain BSM phenomena including Lorentz violation and neutrino decay. An extragalactic component of high-energy astrophysical neutrinos was detected by the IceCube Neutrino Observatory~\cite{Aartsen:2013bka}, but the sources of these neutrinos is still a mystery. The first candidate astrophysical neutrino source, TXS 0506+056, a very-high-energy Blazar, was announced recently~\cite{IceCube:2018dnn,IceCube:2018cha}.
There is also a known yet undetected flux of ultra-high energy neutrinos that are the byproducts of ultra-high energy cosmic ray interactions off background photon fields~\cite{Greisen:1966jv,Zatsepin:1966jv}.
\item{\bf Supernovae:} Neutrinos have been detected from one Supernova~\cite{Hirata:1987hu,Bratton:1988ww}, SN1987A, which provided strong constraints on a great number of BSM phenomena including secret neutrino interactions, tachyonic nature, neutrino decay, Lorentz violation, equivalence principle violations, and others~\cite{Arnett:1987iz,Goldman:1987fg,Manohar:1987ec,Lattimer:1988mf,Raffelt:1987yt,Turner:1987by}. The detection of neutrinos from future galactic supernovae will yield larger samples, better precision, and a different baseline will allow for further enhancements to these constraints. When coupled with an improved understanding of supernova explosion mechanisms, the effects of novel neutrino interactions or exotic particle production on the explosion itself is also constraining~\cite{Arguelles:2016uwb,Bar:2019ifz,Sung:2019xie}. Finally, it may be possible in the next generation of experiments to detect the diffuse supernova neutrino background, which could offer new opportunities to probe new physics~\cite{Horiuchi:2008jz}.
\item{\bf Solar Atmospheric:} Cosmic rays impinging upon the back of the sun produce showers that create neutrinos \cite{Arguelles:2017eao,Ng:2017aur,Edsjo:2017kjk}.  The very-low density environment, relative to the Earth's atmosphere, ensures that essentially all pions and kaons decay rather than interacting hadronically, producing a harder flux.  At energies of interest to neutrino telescopes, only the neutrinos from the solar limb escape without absorption.  This standard model neutrino production mechanism implies that the sun is in fact a high-energy neutrino point source.  These neutrinos provide a neutrino floor to experiments searching for WIMP annihilation or boosted dark matter produced in the solar core, and are likely to come into reach by the next generation of experiments.
\item{\bf Big Bang:} One second after the Big Bang, neutrinos 
decoupled from matter forming the Cosmic Neutrino Background (C$\nu$B). These relic neutrinos form a diffuse cosmic background with a temperature of $\sim$ 1.95K and a density of around 300 cm$^{-3}$. Cosmological models and BSM physics impact both the C$\nu$B and the Cosmic Microwave Background (CMB) measurements~\cite{Green:2019glg}. While there is very strong indirect evidence of their existence, the C$\nu$B has never been directly observed. 
\end{enumerate}

\subsection{Detector Technologies and BSM Signatures}

\begin{figure}[!htbp]
    \centering
    \includegraphics[width=0.44\linewidth]{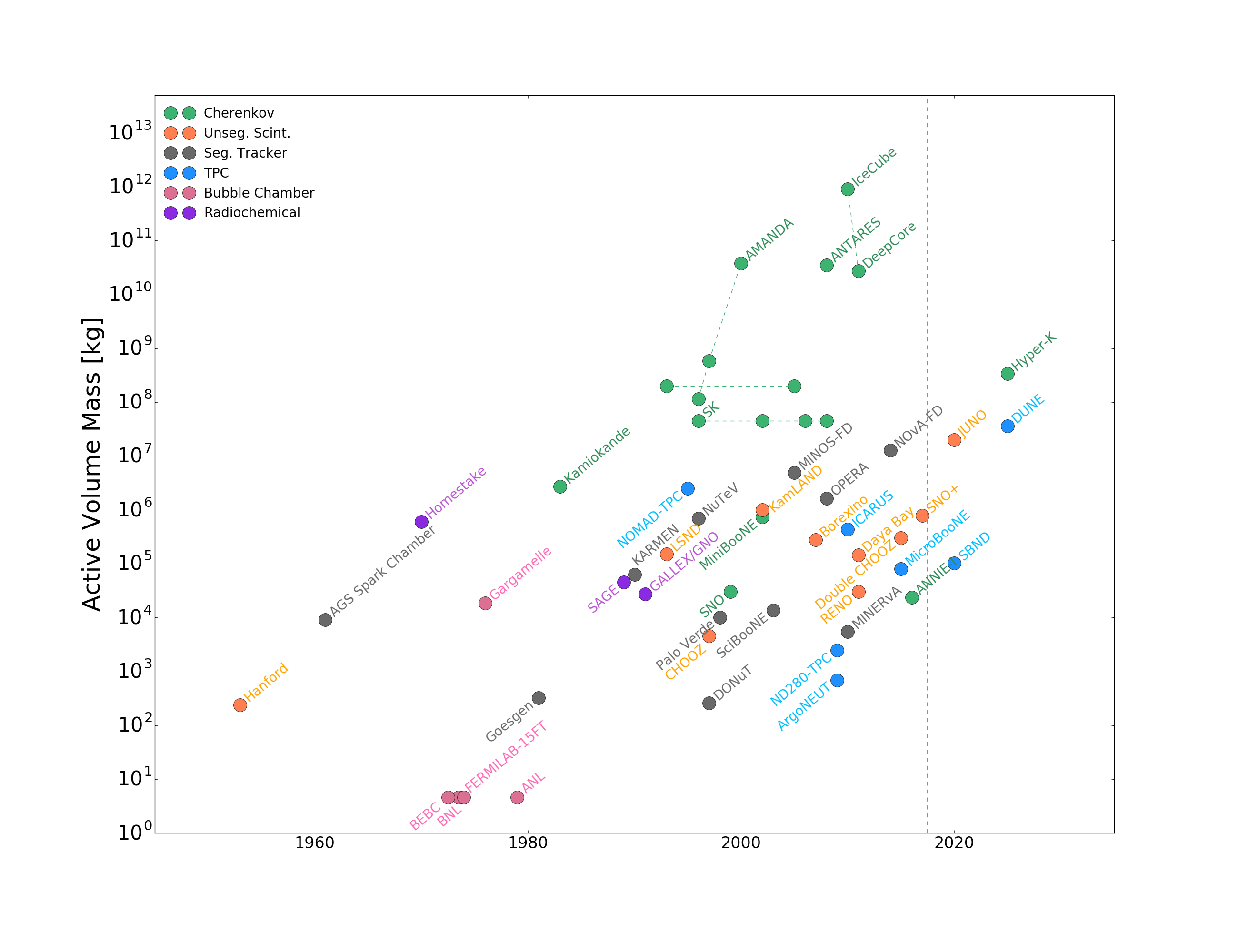}
    \includegraphics[width=0.44\linewidth]{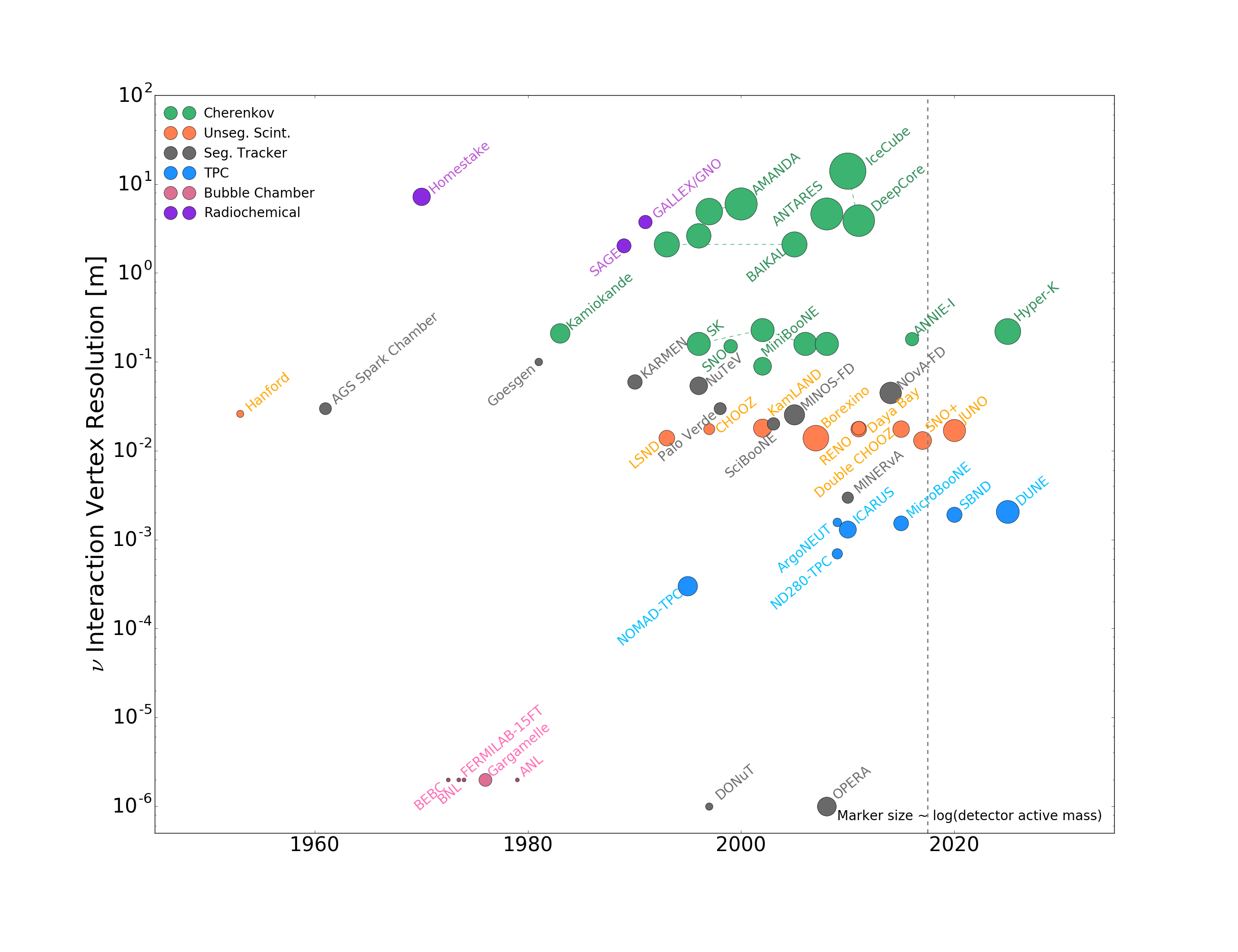}
    \caption{Both of these plots show neutrino detectors as a function of the time of operation in the horizontal axis. These illustrate the progression of our field in two important quantities: active detector mass (left figure) and vertex resolution (right right)~\cite{experiment_plots}. In these figures the color indicates the detector technology. Regarding the left figure. Cherenkov detectors, in water or ice, have achieved the largest active volumes. Unsegmented liquid scintillators have steadily increased in size; the next iteration of this technology, JUNO, is expected to be more than ten times larger than its predecessors. Liquid argon time projection chambers (TPCs) have increased significantly in size, since R\&D efforts in the mid 2010s and in fact the next-generation experiment, DUNE, is expected to be larger than the largest scintillator detector. Regarding the right figure. Different detector technology clearly group themselves in the vertex resolution, here serves as a proxy of fine-grain reconstruction. The emulsion detectors, used {\it e.g.} in OPERA, have the best vertex resolution in the micrometer scale. These are followed by the LarTPC which achieve a resolution of order micrometers. Then the unsegmented liquid scintillator detectors achieve a resolution of order cm. Finally, the water and ice Cherenkov detector have the poorest vertex resolution ranging from 10 cm to 10 meters depending on the detector sparseness.}
    \label{fig:neutrino_detectors}
\end{figure}

The detector technologies and designs planned for the next generation neutrino experiments are chosen to fulfill the primary purpose of an experiment, typically a guaranteed physics return. In Fig.~\ref{fig:neutrino_detectors} we show the evolution of neutrino detectors in both active mass and vertex resolution. 
For example, long-baseline oscillation experiments aim to detect $\nu_e$ signatures in the range of a few 100 MeV to a few GeV, reactor experiments aim to detect few MeV $\bar{\nu}_e$, astrophysical neutrino observatories aim to detect high-energy neutrino signatures, and so forth. The advanced capabilities of the next generation of detectors can also open up sensitivity to BSM physics or dark matter signatures beyond the primary purpose of the experiment. A short survey of the different detector technologies and capabilities deployed in the next generation of neutrino experiments and their impact on sensitivity to BSM signatures is presented.

\subsubsection{Scintillator Detectors}
Scintillation technologies have been used in HEP experiments for a long time and therefore are well-understood technologies.  
This section summarizes experiments utilizing scintillation technologies.

{\bf JUNO} is a follow-up to the Daya Bay Reactor Neutrino Experiment, consists of a large acrylic sphere detector filled with 20 kton of liquid scintillator and surrounded by 20$''$ PMTs with the possibility of using smaller, 3$''$ PMTs to fill in the gaps. This yields an excellent energy resolution of $\sim 3\%/\sqrt{E}$. The detector will be situated approximately 700 m deep underground and submerged in a cylindrical water pool instrumented with 20'' PMTs and top scintillator trackers for accurate cosmic-muon reconstruction and veto. JUNO will measure oscillation rates of reactor anti-neutrinos after travelling approximately 53 km. Besides the precision neutrino oscillation measurement, including determining the neutrino mass ordering, JUNO also has a large BSM program in solar and atmospheric neutrinos, supernova burst and diffuse supernova neutrinos, nucleon decay, dark matter, and non-standard interactions. Details can be found in~\cite{An:2015jdp}.


{\bf Theia} is a proposed very-low threshold novel scintillator detector for long-baseline neutrino oscillation, double-beta decay, solar, geoneutrino, and supernova neutrino studies~\cite{Gann:2015fba}.
Theia is a large water-based liquid scintillator (WbLS) detector.
WbLS is a cost effective, highly tunable, novel medium benefiting from the characteristics of both scintillator and water Cherenkov detectors.
It allows the detection of charged particles below the Cherenkov threshold in a 100 kton scale detector at a lower cost compared to oil scintillator detectors.
The base design for the Theia target is a 50 kton volume instrumented with more than 100,000 photosensors, which include conventional PMTs and large-area picosecond photoDetectors (LAPPD),  in order to reach an effective photocoverage of more than 90\%.
The use of the very-fast LAPPDs enable the large photo-coverage with very fast timing; in the 10ps scale.
In addition to the lower detection thresholds achievable, Theia would enable improved background reduction, which will be specially useful for proton decay or solar neutrinos studies, due to improved photo-detector timing and utilisation of below-Cherenkov threshold scintillation light.
Placing Theia in the Sanford Underground Research Facility (SURF), in addition to the DUNE detectors, would significantly expand on the physics reach of the DUNE LArTPC detectors.

{\bf Very-Short Baseline Reactor Neutrino Detectors}
There is an ongoing world-wide program of very short baseline ($\sim$10 m) reactor neutrino experiments in search of sterile neutrino oscillations
\cite{Vogel:2015wua,Qian:2018wid} including DANSS \cite{Alekseev:2016llm}, NEOS \cite{Ko:2016owz}, PROSPECT \cite{Ashenfelter:2018jrx}, SoLid \cite{Abreu:2017bpe}, STEREO \cite{Allemandou:2018vwb}, Neutrino-4 \cite{Serebrov:2018vdw}, and others. Almost all of these experiments use liquid or solid scintillator detectors at shallow depth constrained by the locations of the reactor complex. Most detectors are doped with Gd or $^{6}$Li to increase the capture cross section of the delayed
neutrons and improve pulse shape discrimination to reject backgrounds. In
addition to sterile neutrino searches, these experiments will provide precision and high-resolution reactor neutrino spectrum measurements, mostly from highly-enriched $^{235}$U research reactors, which will largely reduce the flux uncertainty in other reactor experiments such as JUNO and the reactor CE$\nu$NS program. It is anticipated that these experiments will become systematics limited quite quickly, over the next few years, due to the incredibly high flux at detectors very close to reactors.

\subsubsection{Gas Detectors}
The DUNE near detector complex is proposed to include a high-pressure, approximately 10 bar, argon methane mixture TPC~\cite{Martin-Albo:2016tfh}.  This may in principle give access to novel BSM signatures beyond those accessible with liquid argon detectors, as the lower density enables a more precise measurement of vertex substructure in neutrino scattering, which may also enable detection of electromagnetic BSM signatures which would be difficult to detect in liquid argon.  The gas detector also allows improved reconstruction of BSM signatures such as tridents, and the inclusion of a magnetic field dramatically improves capability by sign-selecting muons and other charged particles.  Finally, a gas detector will enhance sensitivity to BSM signatures by providing detailed characterizations of neutrino interactions and more precisely constraining all components of the neutrino flux.

\subsubsection{Noble Liquid Detectors}

The next generation neutrino experiments utilizing noble liquid detectors are dominated by the Liquid-Argon Time-Projection Chambers (LArTPC) of the Short-Baseline Neutrino program (SBN) at the Fermilab Booster Neutrino Beam (BNB) and the Deep Underground Neutrino Experiment (DUNE) detectors. 

{\bf The Booster Neutrino Beam Neutrino Detectors:}
\begin{figure}[!htpb]
    \centering
    \includegraphics[width=0.6\textwidth]{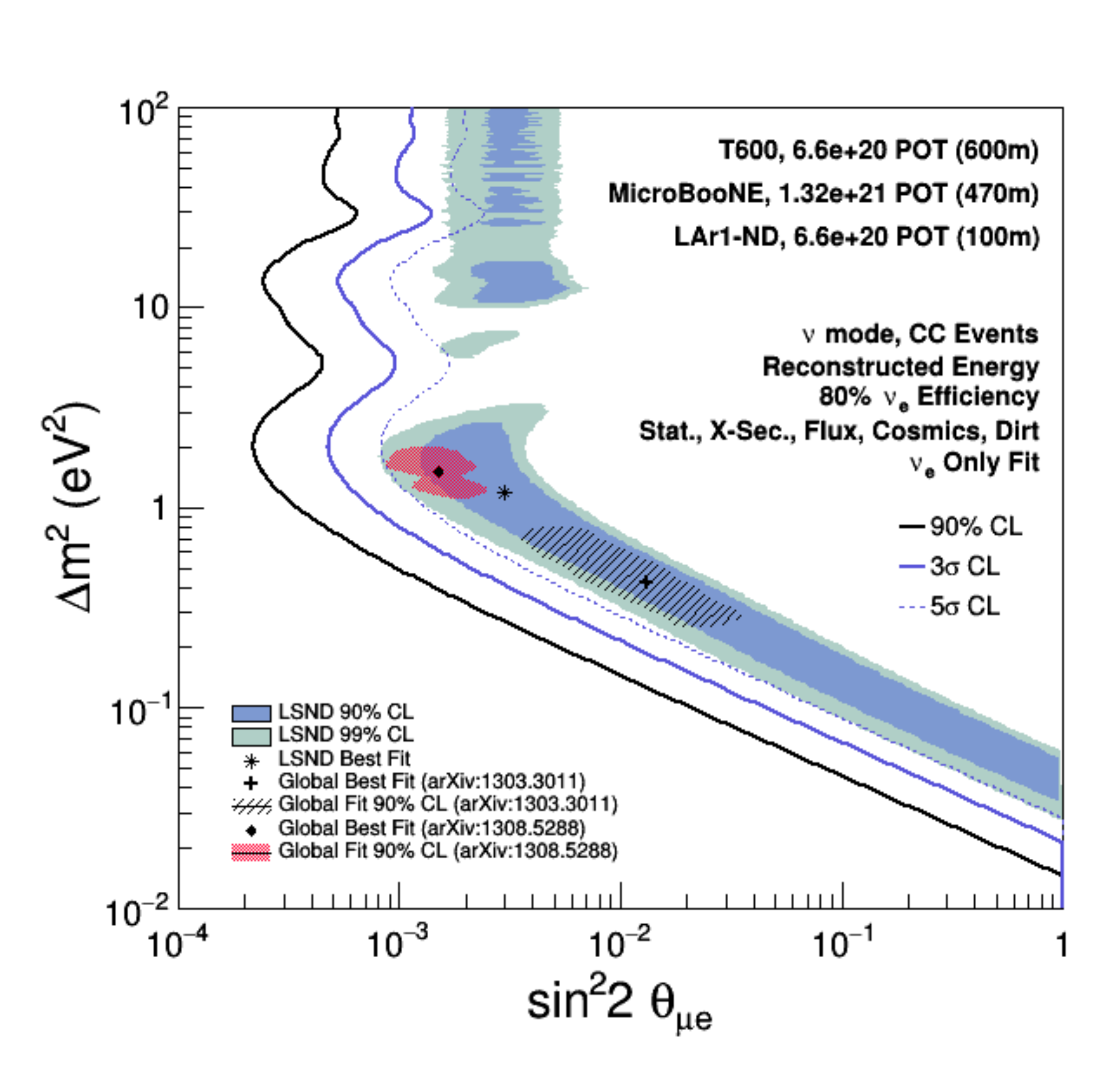}
    \caption{SBN program sensitivity to short-baseline neutrino oscillations in $\Delta m^2$ vs $\sin^2 2 \theta_{\mu e}$ at 90\% C.L. is shown as a solid black line. Two sets of global-fit confidence regions are shown for reference.}
    \label{fig:sbnsens}
\end{figure}

The SBN program~\cite{Antonello:2015lea} at FermiLab comprises of three single-phase LarTPC detectors located at different distances on-axis from the 8 GeV Booster Neutrino Beamline (BNB) as summarized in Tbl.~\ref{tab:sbn}. 

\begin{table}[!htp]
    \centering
    \begin{tabular}{l|cc}
         Detector & Distance from BNB target & Mass \\ \hline
         ANNIE & 100 m (off-axis) & 26 ton Gd-loaded Water total \\
         SBND $^{*}$ & 110 m & 112 ton LAr active\\
         MicroBooNE & 470 m & 87 ton LAr active\\
         MiniBooNE & 541 m & 818 ton Mineral Oil total \\
         ICARUS-T600$^*$ & 600 m & 476 ton LAr active \\
    \end{tabular}
    \caption{The location and active masses of the current and under construction ($^{*}$) neutrino detectors located on-axis to the Booster Neutrino Beam at FermiLab. The ANNIE detector, which aims to study neutron interactions, is also indicated.}
    \label{tab:sbn}
\end{table}
The primary goals of the SBN program is a high-precision search for sterile neutrinos and providing a test bed for LArTPC technology.
The sterile neutrino oscillation program uses $\nu_\mu \rightarrow \nu_{\mu,e}$ oscillations in a wide range of $L/E$ using multiple detectors at different baselines ranging from 100 to 600m with neutrino energies of 200 MeV to 2 GeV. Figure~\ref{fig:sbnsens} summarizes the sterile sensitivity of the SBN program in the $\nu_e$ appearance channel. The MicroBooNE detector is the first LArTPC detector built in the SBN program. MicroBooNE has been operating since 2015 with the goal of determining the nature of the excess of $\nu_e$-like events observed in the larger MiniBooNE Cherenkov detector~\cite{Aguilar-Arevalo:2018gpe}. 

The promise of much improved $e/\gamma$ separation and higher $\nu_e$ reconstruction efficiency expected using the LArTPC technology will enable MicroBooNE to determine whether the MiniBooNE excess is due to $\nu_e$ or other backgrounds. The MicroBooNE collaboration is currently refining the LArTPC simulation and neutrino reconstruction techniques for the low-energy excess analysis. The ICARUS detector has been refurbished and installed at Fermilab and is expected to start collecting BNB data by the end of 2019. SBND is still under construction, but data is expected with all three LArTPC detectors operational by late 2020. 

There are several BSM searches envisioned using the detectors deployed in the Booster Neutrino Beamline:
\begin{enumerate}
    \item {\it Sterile neutrino searches using multiple detectors at different baselines:} Sterile neutrinos can be detected by looking for anomalies in $\nu_\mu$ CC disappearance signatures, $\nu_e$ CC appearance signatures and $\nu$ NC disappearance. This is the primary physics program of SBN; see Fig.~\ref{fig:sbnsens}.
    \item {\it Searches for dark matter and heavy leptons produced in the BNB target:} BSM searches in the SBN detectors at different distances from the target are planned. These searches are limited by the neutrino background and the surface location of the slow LArTPC detectors, which implies that these detectors have a significant cosmic backgrounds. All the LArTPC detectors are equipped with photon sensors in the TPC that enables time tagging of interactions when matched to the ionization clusters reconstructed in the TPC. In addition ICARUS is also equipped with a Cosmic Ray Tagger (CRT) system to further reduce cosmogenic backgrounds. A beam off-target run with the ICARUS detector similar to the MiniBooNE special run~\cite{Aguilar-Arevalo:2018wea} could be attempted after the SBN program has reached its goals. It should be noted that the ANNIE detector currently deployed off-axis in the BNB is primarily focused on measuring neutrons in neutrino-nucleus interactions. The detector is quite small, but in later phases is expected to be instrumented with very-fast LAPPD photon detectors (10~ps resolution). The results from ANNIE could open up an opportunity to use a fast off-axis detector to search for DM and BSM signatures produced in the BNB target or proton dump which could take advantage of the timing structure of the short BNB pulse. 
\end{enumerate}
{\bf Deep Underground Neutrino Experiment (DUNE) Detectors:}
The DUNE experiment~\cite{Abi:2018rgm,Abi:2018dnh,Abi:2018alz} is a long-baseline neutrino experiment located on the future LBNF neutrino beamline from the FermiLab 120 GeV Main Injector. The experiment will contain both a near and far detector complex. The far detector (FD) is envisioned to be 4 LArTPC modules each with 10 kton fiducial LAr. The FD is located 1297~km from the LBNF target in the Sanford Underground Research Facility in Lead, South Dakota. Currently two LArTPC technologies are envisioned for the FD modules: a single-phase LArTPC utilizing projective wire planes and cryogenic readout electronics in the LAr; and a dual-phase LArTPC which detects the ionization charge in the gas phase above the LAr using Large Electron Multipliers (LEMS) that amplify the electrons in avalanches that occur in the gas phase. The DUNE experiment is currently projected to start taking data with the first 10 kton single-phase module around $\approx$ 2027.
The near detector (ND) complex plans to include an unmagnetized single-phase LArTPC with pixelated readout followed by a magnetized low-density tracker.
The DUNE experimental complex is sensitive to BSM physics through different approaches: 
\begin{enumerate}
\item  {\it Direct BSM searches in the near detector complex} similar to those in the SBN LAr detectors: For example heavy neutral lepton (HNL) production from decays of mesons produced in the target. The HNL travel to the ND and the cleanest signature are decays to two charged particles - like $N \rightarrow \mu^+ \pi^-$. The DUNE ND complex is still being defined, but an unmagnetized LArTPC with pixelated readout is a core component. The pixelization improves the performance of the DUNE ND compared to the existing projective wire plane readout used in the SBN program detectors and addresses the high rate of neutrino interactions expected in the DUNE ND LArTPC. Improved vertex resolution and $e\gamma$ separation is expected. The pixelated readout also addresses loss of tracking efficiency and $dE/dX$ resolution for tracks at angles aligned with one of the projective wire planes in the standard designs. Tracks perpendicular to the wire planes produce prolonged signals on the induction planes which can be difficult to extract. Exact performance numbers are not yet available, and recent advances in signal processing and 3-D imaging using projective wire readout ~\cite{Qian:2018qbv} has potentially narrowed the performance gap. 

In the current DUNE ND complex the unmagnetized LArTPC is complemented by a highly capable magnetized low-density tracker - currently realized by a gas Ar TPC in a magnetized volume. This enables the DUNE ND complex to achieve better momentum resolution for charged tracks produced in the LArTPC target volume but which enter the magnetized high precision tracker. A proposal is currently underway to move several components of the DUNE ND  - including the LArTPC - off-axis up to $\sim 3.5^{\circ}$ (DUNEPrism) to enable measurements of neutrino fluxes with different energies. Improved sensitivities to BSM signatures would be possible in the off-axis locations due to the lower neutrino background ~\cite{DeRomeri:2019kic}.
The higher energy range, better vertex resolution, charge measurements and improved momentum resolution of the DUNE ND complex will extend the range of BSM searches in the LArTPC beyond that achievable by the SBN program. The actual gain will depend on the specific signature. 
\item {\it Indirect BSM searches through precision measurements of neutrino scattering} properties in the ND complex for example through precision measurements of $\sin^2 \theta_W$.
\item {\it BSM searches using the ND as a short-baseline oscillation experiment:} The DUNE ND is located at a distance of 570m from the neutrino beam target, but is illuminated by a neutrino beam with energy peaked at 2.5 GeV (reference beam design) and a beam with 5-10 GeV neutrinos (high energy beam tune option). The $L/E$ range is thus different from the SBN program based on the Booster Neutrino beamline.
\item {\it Long-baseline interferometry:} using precision measurements of long baseline $\nu_\mu \rightarrow \nu_x$ neutrino oscillations, where $x = e,\mu, \tau$ with the 40 kton LAr far detector. The far detector is designed to be sensitive to electron and muon neutrino signatures in the energy range of few 100 MeV up to 10's of GeV. Electrons and photons leave large EM showers in the LArTPCs and are differentiated using $dE/dX$ and topology (gap before photon converstion for example). 

Precision measurements of the shape of the oscillation signal is the key to many BSM searches (NSI, steriles neutrinos, etc) and totally active LArTPC are expected to have both excellent EM energy resolutions (order of few \%/$\sqrt{E}$) and muon momentum resolution from range (order of few \%) for contained muons and ($\sim 10\%$) from multiple-scattering measurements for exiting muons. Hadronic energy resolution is of the order $30\%/\sqrt{E}$. Protons are easily identified and measured through range and $dE/dX$. $\nu_\tau$ signatures are detected using $\tau$ decay to electron, muons and hadrons and are similar to $\nu_\mu, \nu_e$ CC and $\nu$ NC signatures but with slightly different event topologies and energy distributions. While neutron detection is somewhat possible in LArTPCs  - through hadronic interactions or scattered protons - it is difficult to associate neutron signatures to the original interaction. 
\item {\it Searches for nucleon decay:} both proton decay and $n-\bar{n}$ oscillations. The most promising proton decay signatures are based on modes with charged kaons in the final state with energy ranges in range of a few 10s-100s MeV. The kaons are identified by range and $dE/dX$ followed by its signature decays. The largest background to the nucleon decay searchs in the DUNE LArTPCs are atmospheric neutrinos.
\item {\it Direct searches for non-accelerator BSM signatures in the far detector:} A representative example would be searches for boosted dark matter with mass in the range of a few to 10's of GeV, the characteristic signals of which are recoil electrons and protons. 
\end{enumerate}

\subsubsection{Water Cherenkov Detectors}
{\bf Super-K and Hyper-K}
Super-Kamiokande (Super-K, SK) is the world's largest underground water Cherenkov experiment, comprising of 50 kt volume, out of which 22.5 kt is the fiducial volume. It is located beneath a one-km rock overburden, approximately 2700 meter-water-equivalent, within the Kamioka mine in Japan.
The SK detector is composed of an inner (11,146 inward-facing 20-inch PMTs, providing 40\% photo-coverage) and an outer (1,855 8-inch outward-facing PMTs) detector, which are optically
separated. Cherenkov radiation produced by charged particles traveling through water is collected by the PMTs and is used to reconstruct the events. The detector has energy threshold of $\sim5$ MeV.
Data collected by Super-Kamiokande (SK) during the run periods of SK-I to SK-IV, starting in 1996, corresponds to a combined exposure of more than 350 kiloton-years. Details of the detector design, performance, calibration, data reduction, and simulation can be found in~\cite{Fukuda:2002uc,Abe:2013gga}. 

Among many other advances, Super-Kamiokande provide definate proof of flavor conversion in atmospheric neutrinos~\cite{Fukuda:1998mi}.
Due to its very general detection capabilities, the experiment allows one to study an expansive variety of physics phenomena, ranging MeV to sub-TeV in energies. These include leading results using solar and atmospheric neutrinos (oscillations parameters measurements~\cite{Abe:2016nxk,Abe:2017aap}, tests of Lorentz
invariance~\cite{Abe:2014wla}, first evidence for day-night asymmetry and terrestrial neutrino matter-effects~\cite{Renshaw:2013dzu}, neutrino magnetic moment tests~\cite{Liu:2004ny}, sterile neutrino searches~\cite{Abe:2014gda}, indirect dark matter searches~\cite{Choi:2015ara,Kachulis:2017nci}), supernovae
relic neutrino searches~\cite{Zhang:2013tua}, exotics searches~\cite{Ueno:2012md,Takenaga:2006nr}, as well as probes of baryon number violating processes~\cite{Takhistov:2016eqm} such as nucleon decays ({\it e.g.}~\cite{Miura:2016krn}), di-nucleon decays and neutron-anti-neutron ($n-\overline{n}$) oscillations~\cite{Abe:2011ky}. 

The Super-K detector has recently undergone a successful refurbishment. Very soon (2019 planned)  gadolinium will be dissolved in water~\cite{Beacom:2003nk} and the SK-Gadolinium (SK-GD) experiment will start. The upgrade will allow neutron tagging with  $\sim 90\%$ efficiency. This will drastically improve sensitivity to supernovae neutrinos through inverse beta decay ($\overline{\nu} + p \rightarrow e^+ + n$) and  allow for further background reduction in proton decay studies. SK-GD is capable of first detection of the diffuse supernova neutrino flux within few years of operation.

{\bf Hyper-Kamiokande (Hyper-K, HK)}~\cite{Abe:2018uyc} is the next-generation water Cherenkov experiment in Japan and successor of the running Super-K. The experimental construction, to begin 2020, will take place in Tochibora mine in Japan. The initial experiment will consist of a single water tank of 260 kiloton (187 kiloton fiducial) volume, surrounded by PMTs with higher detection efficiency than in SK. The experiment will have ten times more statistics than SK and will allow for unprecedented sensitivity to studies of neutrino oscillation, proton decay and astro-particle phenomena. After successful deployment of the first tank, construction of a second detector tank is foreseeable~\cite{Abe:2016ero}.

{\bf T2K and T2HK} uses SK or Hyper-K as a far detector for neutrinos produced from the J-PARC 30 GeV Main Ring located $2^{\circ}$ off-axis and 295~km away. The J-PARC off-axis neutrino beam is narrow-band with a peak energy at 600 MeV. The Super-K and Hyper-K Cherenkov detectors detect the muons and electrons from quasi-elastic beam $\nu_{\mu,e}$ interactions with high efficiency and purity. The shorter baseline of this setup compared to DUNE increases the sensitivity to BSM that manifests in long-baseline oscillations when the results from both experiments are combined. Both experiments cover similar $L/E$ range, although the DUNE experiment has more coverage on the second oscillation maximum. The detectors and neutrino energies detected are very different and have very different systematic uncertainties. T2HK will collect neutrino oscillation data at a much higher rate compared to DUNE due to the larger detector size. 

{\bf ANTARES}
The Astronomy with a Neutrino Telescope and Abyss environmental RESearch (ANTARES) project ~\cite{Zornoza:2012df} is located 2.5~km deep in the Mediterranean Sea and floats 40~km off the coast of Toulon, France. It is composed of twelve 350 meter tall strings, spaced 70 meters apart, that are anchored to the sea floor. The top of each string is held aloft by a buoy. Each string has 75 pressure-resistant digital optical modules (DOMs) containing a single PMT for detecting Cherenkov radiation produced by particles resulting from neutrino interactions. 

While primarily serving as a high-energy neutrino observatory (\SI{10}\TeV\;- \SI{10}\PeV), it is designed to indirectly detect dark matter interactions through the observation of neutrinos produced in WIMP dark matter annihilations via direct processes or Kaluza-Klein scenarios. It currently has great angular resolution for $\nu_{\mu}$ CC events ($\sim$\SI{0.4}\degree) and ($\sim$\num{2}-\SI{15}\degree) resolution for NC and $\nu_{e}$ and $\nu_{\tau}$ CC. However, ANTARES suffers from poor energy resolution ($\sim$100\%), high-absorption rates, and a significant background arising from $^{40}$K and bio-luminescence. Due to the energy resolution constraints, efforts to study neutrino oscillations have been hindered. 

{\bf KM3NeT}
The Cubic Kilometer Neutrino Telescope (KM3NeT)~\cite{Bagley:2009wwa} hopes to expand on the initial design of ANTARES and address several issues encountered. In the new design, detector components will be distributed across three sites off the coasts of France, Greece, and Italy: totalling six hundred mDOM strings. The mDOMs themselves will be expanded to contain 31 PMTs each. Angular and energy resolution for track reconstruction have been improved in the new design to $\sim$\SI{0.2}\degree and \SI{86}\percent, respectively. There is no predicted change in the resolution of cascade reconstruction. 
 
{\bf BDUNT/Baikal-GVD}
The Baikal Deep Underwater Neutrino Telescope (BDUNT), a similarly sized detector as ANTARES, finished deployment in 1998 and has been operating since, with an upgrade in 2005. The experiment's successor, Baikal Gigaton Volume Detector (Baikal-GVD), started deployment in 2015 and currently has three of the planned eight clusters~\cite{Avrorin:2018ijk} deployed. When completed in 2020 it will have an effective volume of 0.4 km$^3$.

{\bf CHIPS}
 is a water Cherenkov detector that uses multiple locations within lakes created by disused mine pits to cheaply create a megaton neutrino oscillation experiment in the line of site of a neutrino beam. Being surrounded by water reduces the need for structural supports while simultaneously providing a cosmic-ray shield. Arrays of PMTs within the detector allows reconstruction of the neutrino events with a higher energy threshold and similar energy resolution to Super-K. This detector is expected to use the NuMi neutrino beam at FNAL and is currently been deployed.

\subsubsection{Ice Detectors}
{\bf IceCube and its upgrades}
The IceCube experiment is a pioneering, cubic-kilometer scale, ice-Cherenkov neutrino telescope at located the South Pole. IceCube's power to constrain BSM physics stems from its vast size, which allows access to fluxes of high-energy atmospheric ($\sim$\SI{5}\GeV\;- \SI{100}\TeV) and astrophysical ($\geq$ \SI{100}\TeV) neutrinos to be measured. IceCube's event sample divides the data into two broad categories: tracks, with precise directional resolution O(\SI{0.1})$^\circ$ but poor energy resolution ($\sim$\SI{50}\percent), and cascades with good deposited energy resolution ($\sim$\SI{50}\percent\;at \SI{10}\GeV, $\sim$\SI{10}\percent\;at \SI{1}\PeV) and poor directional resolution ($\sim$\SI{60}\degree\;at \SI{10}\GeV~and~$\sim$\SI{10}\degree\;at \SI{1}\PeV), which provide two complimentary channels in which to search for new physics. A third potential morphology, double-bangs (also called double cascades), have been explored and only recently detected. Additional tau-neutrino induced morphologies, such as lolipops, inverted lolipops, suggar-daddys, and tautie-pops are yet to be observed. 

IceCube is the only experiment which has convincingly observed a high-energy astrophysical neutrino flux, and this handful of events with uniquely long-propagation distance and high-energy allow constraints to be placed on Lorentz violation, long-range interactions, and other physics that may modify the expectation from standard oscillations. Increasing the size of the detector, as conceived for the future IceCube Generation Two (IceCube-Gen2) program, will dramatically increase the sample size of astrophysical neutrino candidates and enhance capabilities to constrain new physics via measurements of the astrophysical neutrino flavor composition and energy spectrum.

IceCube also has great power to constrain lower energy phenomena using a large statistics atmospheric neutrino sample. Matter-enhanced resonant oscillations of \si{\eV} sterile neutrinos have been explored and constrained in the \si{\TeV} range. At lower energies still, non-standard interaction effects, neutrino decay, and other sterile neutrino signatures have been explored. The deployment of denser arrays, beginning with the IceCube Upgrade, will lower the energy threshold in a section of the detector. The IceCube Upgrade will allow for much tighter control of ice-related systematics, such as the properties of the refrozen hole-ice column and the nature of the anisotropic scattering and absorption within the bulk ice, which both limit low-energy searches. This is because shorter DOM-to-DOM distances and a wide variety of calibration distances will measure the properties of the ice local to the DOMs and constrain models of the bulk ice structure. Improved understanding of the ice properties within IceCube also enhances the capability to search for double-bangs, both from high-energy $\nu_\tau$ but also potentially from neutral heavy leptons and other BSM scenarios at lower energy. Better understanding of ice properties can also enhance the directional resolution for cascades, and can be applied retroactively to data collected by the first generation IceCube detector. This upgrade has been approved and is planned to be completed in the following years.

{\bf Radio}
Experiments pursuing the ultra-high energy (UHE) neutrino frontier are able to probe neutrino physics and BSM scenarios at EeV enrgy scales. Such experiments look for radio showers caused by tau-lepton decays from earth- or mountain-skimming tau neutrinos in the ice or over land. 
This channel is sensitive to extreme non-thermal sources of neutrinos, and can measure the cosmogenic flux.
The cosmogenic flux is a guaranteed flux of UHE neutrinos coming from the interactions of ultra-high energy cosmic rays (UHECRs) with the CMB.
A measurement of the cosmogenic flux will provide orthogonal constraints of UHECRs~\cite{Moller:2018isk,AlvesBatista:2018zui}, but could also be used as unique probes of new physics, in particular in the tau neutrino sector.
Current experiments such as AUGER~\cite{Abraham:2007rj}, ANITA~\cite{Gorham:2016zah}, and ARA~\cite{Allison:2014kha} have set limits on the UHE neutrino flux, and future detectors like GRAND~\cite{Alvarez-Muniz:2018bhp}, RNO (the merger of ARA~\cite{Allison:2011wk} and ARIANNA~\cite{Barwick:2016mxm} efforts), and POEMMA~\cite{Olinto:2017xbi} will likely detect them.

Recently, ANITA reported two anomalous upward-going UHE air showers~\cite{Gorham:2016zah,Gorham:2018ydl} with extensive air shower energies approaching 1 EeV. Standard Model explanations have been ruled out at more than $5 \sigma$ level~\cite{Connolly:2011vc,Chen:2013dza,Albacete:2015zra,Bertone:2018dse,Motloch:2016yic,Bustamante:2017xuy,Romero-Wolf:2018zxt}. However, a wide range of BSM interpretations have been proposed, including heavy dark matter decay~\cite{Anchordoqui:2018ucj,Dudas:2018npp,Heurtier:2019git} and sterile neutrino mixing~\cite{Cherry:2018rxj,Huang:2018als} as well as connections to supersymmetry~\cite{Fox:2018syq,Collins:2018jpg,Chauhan:2018lnq,Anchordoqui:2018ssd}.

\subsubsection{CEvNS detectors}

The Coherent Elastic Neutrino Nucleus (CEvNS) cross-section is a well-known SM process~\cite{Freedman:1973yd}. There is a wide variety of experiments utilizing very different detector technologies and neutrino sources to measure this process on different nuclei in progress and planned. Deviation of the cross-section measurements from expectations are signatures of BSM. In addition, the experiments can also search for other BSM signatures such as sterile neutrinos or NSI. Experiments based at beam dumps are also sensitive to dark matter produced in the target or beam dump.
Experimental progress has recently been rapid by leveraging advances made in the direct detection community combined with powerful sources of neutrinos.

{\bf \coherent}
The \coherent experiment~\cite{Akimov:2018ghi} operates a suite of detectors with its primary goal being observation and characterization of coherent elastic neutrino-nucleus scattering (CEvNS); they reported the first detection of CEvNS in 2017 at 6.7 $\sigma$ significance with respect to the null~\cite{Akimov:2017ade}. The detectors are located about 20--30 m from the target of the Spallation Neutron Source (SNS) at the Oak Ridge National Laboratory. The SNS bombards a mercury target with 1-GeV protons at 1.4 MW which serves as a very clean and intense source of $\pi$-decay-at-rest neutrinos.
The \coherent detectors currently taking CE$\nu$NS data are 14.6-kg CsI and 22-kg LAr detectors, and the collaboration is planning to install a 3388-kg NaI detector, a 612-kg LAr detector, and Ge detectors with the total mass of 16 kg in the next few years. 

Numerous up-coming experiments are also hoping to measure CE$\nu$NS soon, all using reactor neutrinos. For example:

{\bf CONUS}
COherent NeUtrino Scattering experiment (CONUS)~\cite{Hakenmuller:2019ecb}: 4-kg p-type point-contact HPGe detectors 17 m from a 3.9-GW reactor at the Brokdorf nuclear power plant in Germany.
In 2018 CONUS claimed a $\sim2.3$~$\sigma$ evidence for CEvNS consistent with the SM prediction \cite{maneschg_werner_2018_1286927}.
{\bf MINER}
Mitchell Institute Neutrino Experiment at Reactor (MINER)~\cite{Agnolet:2016zir} utilizes SuperCDMS detectors to detect neutrinos produced by a 1-MW nuclear reactor with a possibility of changing the baseline from meters to tens of meters. 
{\bf TEXONO}
Taiwan EXperiment ON neutrinO (TEXONO)~\cite{Soma:2014zgm}: point-contact Ge detectors 28 m from a 2.9-GW reactor at Kuo-Sheng Reactor Neutrino Laboratory in Taiwan.
{\bf CONNIE}
COherent Neutrino Nucleus Interaction Experiment (CONNIE)~\cite{Aguilar-Arevalo:2016qen}: gram-scale Si CCD detectors 30 m from a 4-GW nuclear reactor near Rio de Janeiro, Brazil.

\subsubsection{Neutrinoless Double Beta Decay Detectors and Direct Dark Matter Searches}
Neutrinoless Double Beta Decay and Dark Matter experiments were outside the scope of this work. However, the central mission of these experiments involves searching for BSM phenomena related to dark matter and neutrinos, and their precision detection techniques obviously allow for sensitivity to other BSM interactions. 

The needs of direct detection dark matter searches and $0\nu\beta\beta$ experiments are distinct from one another, although there is much technological overlap between them. The primary goal of both experiments is the minimization of backgrounds in the energy range of interest. For $0\nu\beta\beta$, this is a mono-energetic peak comprised of two electrons around the Q-value of the isotope of interest, a few MeV. Thus experiments with good energy resolution, topological capabilities, and potentially also daughter-ion tagging capabilities are required.  Next-generation neutrinoless double-beta decay proposals include ton-scale or larger arrays of germanium diodes (LEGEND~\cite{Abgrall:2017syy}), high pressure xenon gas (NEXT~\cite{Martin-Albo:2015rhw}), liquid xenon (nEXO~\cite{Albert:2017hjq}, Darwin~\cite{Aalbers:2016jon}), and scintillating bolometers (CUORE~\cite{Alduino:2017ehq}, CUPID~\cite{Wang:2015raa}).  For dark matter experiments, the goal is detection of very-low energy events, implying a distinct suite of backgrounds and somewhat different detection needs including low thresholds, electron vs nuclear recoil discrimination (for example via pulse shape discrimination).  Next-generation technologies include liquid xenon (LZ~\cite{Akerib:2015cja}, XenonNT~\cite{Aprile:2014zvw}, Darwin), liquid argon (DarkSide~\cite{Agnes:2014bvk}), bubble chambers (PICO~\cite{Amole:2017dex}), and silicon and germanium detectors (SuperCDMS~\cite{Agnese:2014aze}). 

Dark matter and double beta decay experiments may also have sensitivity to exotic Majoron emitting decay modes, millicharges, nucleon decays and other BSM phenomena.

\subsection{Radioactive Source Experiments}

There are several experiments that study neutrinos from radioactive sources. Two technologically very different experiments are discussed below as examples of the potential for BSM searches using neutrinos from radioactive sources:

{\bf PTOLEMY:} The Princeton Tritium Observatory for Light, Early-Universe, Massive-Neutrino Yield (PTOLEMY) experiment proposes to search for relic Big-Bang neutrinos through capture on a $\beta$-decaying nucleus, in this case Tritium. The signature of BB neutrinos are electrons that are clustered above the $\beta$-decay endpoint. The Tritium source is challenging since it requires a 100~g of Tritium trapped in a single atomic layer of Graphene to maximize surface area. The electron capture and spectrometry techniques are similar to Tritium end-point experiments targeted at measuring the neutrino mass through accurate measurements of the $\beta$-decay end point. In PTOLEMY sensitivity to low-mass sterile neutrinos that mix with electron neutrinos would manifest as a narrow peak of electrons at an energy equivalent to the end point energy plus the mass of the heaviest mass eigenstate. Directional detection of MeV Dark Matter may also be possible with PTOLEMY.

{\bf HUNTER:}  The Heavy Unseen Neutrinos from Total Energy-momentum Reconstruction (HUNTER) experiment proposes to search for keV sterile neutrinos by energy-momentum reconstruction of atomic K-capture events. The HUNTER experiment uses cloud of decaying $^{131}$Cs atoms in an atomic trap. The $^{131}$Cs absorbs a K-shell
atomic electron and emits a neutrino, the nucleus recoils to conserve momentum and is followed by emission of an atomic X-ray and one or two
Auger electrons as it de-excites. By measuring accurately both the recoil energy of the nucleus and the energy of the X-rays and 
Auger electrons the neutrino invariant mass can be reconstructed. Reconstructed neutrino masses in the keV range would constitute a signal for sterile neutrinos. 

\subsection{Other Beam Dump and Related High-Intensity Opportunities}

Over the past two years, CERN has been investigating how to upgrade its existing accelerator park for high-intensity beams and 
a beam dump facility for searches of BSM physics in the 
low-mass low-coupling sector. 
A study was launched in 2016 to prepare for the European Strategy for Particle Physics (ESPP) meeting that takes place in  2019-2020, called the Physics Beyond Colliders (PBC) initiative.
The findings are summarized in~\cite{Beacham:2019nyx}.

The PBC initiative is an exploratory study aimed at exploiting the full scientific potential of the CERN’s accelerator complex and scientific infrastructures through projects complementary to the LHC and other possible future colliders. These projects will target fundamental physics questions in modern particle physics. The document presents the status of the proposals presented in the framework of the Beyond Standard Model physics working group, and explores their physics reach and the impact that CERN could have in the next 10-20 years on the international landscape.

About 15 proposals are discussed for experiments searching for axions or axion-like particles, dark photons, dark scalars, millicharge particles and heavy neutral leptons. A beam dump facility is proposed at the 400 GeV SPS accelerator which could
accommodate several experiments, the most outspoken one to be the SHiP experiment, which is proposed to receive $2\times 10^{20}$ protons on target over a time span of 5 years. 

In this document sensitivity plots are reported  that show the reach in couplings versus mass (typically below 10 GeV) for the different proposals and are compared to the already experimentally established limits. Certainly in the longer term the study proposed in this document should take into account these sensitivities and, where relevant,  compare with the potential of the neutrino near detector proposals.



\section{BSM Physics Prospects at the Next-Generation Neutrino Experiments}
\label{sec:exec-prospects}

Current and next generation neutrino experiments will have unprecedented capabilities, ranging from the superb event reconstruction of liquid argon time projection chambers~\cite{Antonello:2015lea, Abi:2018dnh}, to the large oscillation data sets from water or ice Cherenkov~\cite{Abe:2015zbg,TheIceCube-Gen2:2016cap} and liquid scintillator~\cite{An:2015jdp} detectors, the unparalleled probe of very high energy neutrinos in Antarctic observatories~\cite{Aartsen:2014njlMartineau-Huynh:2015hae, vanSanten:2017chb}, and beyond.
In this section, a compendium of several beyond standard model scenarios that can be probed in current and future neutrino experiments is presented. 

\subsection{BSM Physics Tools and Prospect for Future Improvements}

In order to accurately assess the capabilities of current and future neutrino experiments to discover BSM physics and to develop appropriate analysis strategies, accurate simulation of both SM and BSM phenomenology is required. This proves challenging, particularly in the energy regime most relevant to accelerator and atmospheric neutrino experiments, where strongly coupled QCD effects are present.  Simulation of the beam-generated fluxes, non-standard-neutrino and dark matter interactions, and atmospheric neutrino backgrounds are all subject to large uncertainties. In this section, we provide an overview of the current status of tools for simulating these phenomena, discuss areas that are challenging, and assess the potential for future improvement.
\subsubsection{Beam-generated Fluxes}
While accelerator neutrinos are typically produced in the decays of mesons and muons, dark sector states can be produced either in meson decays or directly from the quarks through Drell-Yan process.  
In the case that they are produced in the decays of mesons, standard neutrino flux generation tools such as \verb+g4bnb+ (Booster)~\cite{AguilarArevalo:2008yp} and \verb+g4numi+ (NuMI)~\cite{Aliaga:2016oaz} can be adapted to have mesons decay to dark sector particles instead of neutrinos.
For direct production, simulations can be done directly using collider-oriented codes like 
\verb+MadGraph+~\cite{Alwall:2014hca}
and \verb+Pythia+~\cite{Sjostrand:2014zea}, 
but dedicated codes such as
\verb+BdNMC+~\cite{deNiverville:2016rqh} and 
\verb+MadDump+~\cite{Buonocore:2018xjk} are of greater interest.  In both cases, simulation is subject predominantly to the same hadronic and beam uncertainties that plague neutrino flux simulations, so improvements in understanding those will lead to commensurate improvements in understanding the beam-generated fluxes of dark sector particles.
In this regard, it is worth noting that the Drell-Yan production process becomes meaningless below the dark sector mediator mass of about $~3GeV$ due to these uncertainties. 
Models based on other mediators should also be explored as soon as possible and should be implementable quickly based on existing tools.
\subsubsection{Dark Sector Interactions}
\label{sec:dsint}
Dark matter and dark sector particles produced either astrophysically or in the beam can be detected by their interactions in the active volume of a neutrino detector.  Upcoming neutrino detectors, particularly those based on LArTPC technology, will have unprecedented sensitivity to these interactions in the theoretically challenging region of energy deposits in the 10 MeV to few GeV range.  Nuclear and hadronic modeling are required in this regime, with much of the physics being understood only empirically or based on ad hoc models, rather than first principles.  Neutrino Monte Carlo codes, such as \verb+GENIE+~\cite{Andreopoulos:2009rq,Andreopoulos:2015wxa}, \verb+NEUT+~\cite{Hayato:2009zz}, \verb+NuWRO+~\cite{Golan:2012wx}, and \verb+GiBUU+~\cite{Buss:2011mx} already incorporate modeling of this physics.  The implementation of dark matter interactions in these codes will be of increasing importance given the novel capabilities of upcoming detectors.

To date, dark matter interactions within a vector-mediated model have been implemented~\cite{Berger:2018urf} in the \verb+GENIE+ Monte Carlo used by a large number of experimental collaborations. This widespread use allows it to easily be connected to detector simulation frameworks. The implementation is mostly complete, but simulation of baryon resonance excitation is in progress. Further models, such as inelastic boosted dark matter, should be implemented as soon as possible as well in order to set a broad target for BSM searches at upcoming experiments.

Beyond the usual uncertainties with studying rare interactions with nuclei that affect neutrino Monte Carlo, several new challenges affect dark matter simulation, as no dark matter interactions with nuclei have been observed to date. The quark-level structure of interactions is not known and must be left as a free parameter in modeling. This leads to the introduction of form factors and baryon resonance helicity amplitudes that have not and cannot be studied in data based on standard interactions. Improved studies of these objects on the lattice could help (see Refs.~\cite{Alexandrou:2017hac,Capitani:2017qpc,Rajan:2017lxk,Yao:2017fym,Suzuki:2017ifu,Hasan:2017wwt,Chang:2017eiq,Bar:2018xyi,Hasan:2019noy} for a selection of recent studies of the relevant form factors). More recent nuclear models for baryon resonances could also be interesting to explore~\cite{Leitner:2008ue,Buss:2011mx}.
\subsubsection{Machine Learning at Next Generation Experiments}

A key element in data analysis is developing selection criteria to separate signal and background.  In their simplest form, selection criteria are a collection of thresholds on quantities derived from data which have different characteristics between signal and background.  These are known as {\em rectangular cuts}.  To achieve higher efficiencies and purities than rectangular cuts can achieve, over the last 20 years, High Energy Physics has increasingly turned to machine learning techniques~\cite{Radovic:2018dip}.

The success of convolutional neural networks (CNNs)
has led to rapid progress in high energy physics applications. The first published use of CNNs in high energy physics improved the selection efficiency of $\nu_{e}$ interactions at the NOvA experiment.  The statistical effect of using CNNs was equivalent to increasing the detector mass by 30\%~\cite{NovaCVN}.  Due to this tremendous success, this network was adopted by all NOvA analyses. MicroBooNE used similar techniques to separate cosmic rays from neutrino interactions~\cite{uBooNECNN}.  

In future experiments, deep learning methods show promise for reconstruction, in addition to selection.  MicroBooNE has recently implemented {\em semantic segmentation}, a technique which uses the series of convolutional filters used for categorization to make predictions about what particle produced each energy deposit~\cite{uBooNESemSeg}.  They use this to identify all the reconstructed hits likely produced by electromagnetic particles.  
The HEP.TrkX collaboration has demonstrated using graph neural networks (GNNs) to successfully build tracks in a generic detector~\cite{heptrkX2018}.  GNNs generalize the concept of a CNN to irregular grids by operating on general structures described by nodes and edges.  Taking energy deposits in three dimensions as nodes and the collection of plausible causal connections between them as edges, GNNs should be able to infer the true connects between energy deposits without having to also store large numbers of empty pixels.  GNNs have been used by the IceCube collaboration to improve event classification~\cite{IceCubeGNN}.

\subsection{BSM Physics with Neutrinos}
The neutrino sector is the least known sector of the standard model. 
The mechanism behind neutrino masses and mixings is still an open question.
Therefore, it is natural to view the neutrino sector as a window to physics beyond the standard model.
New physics in the neutrino sector can manifest itself in many distinctive ways.
For instance, the presence of sterile neutrinos may lead to modification of oscillation probabilities, while non-standard neutrino interactions may reveal themselves in oscillation and scattering measurements. 
Light new degrees of freedom may enhance rare neutrino processes such as trident events, or provide novel experimental signatures that can be searched with current and future experiments.
In this section, a summary of several BSM physics scenarios in the neutrino sector and their corresponding phenomenology is presented. 

\subsubsection{Sterile Neutrinos}
%
Electron-volt scale sterile neutrinos have been motivated by the 
observation of $\bar \nu_e$ appearance in $\bar \nu_\mu$ at LSND~\cite{Aguilar:2001ty} and MiniBooNE~\cite{Aguilar-Arevalo:2018gpe}, which probe oscillations at an $L/E \sim 1 {\rm km/GeV}$. These sterile neutrinos extend the PMNS $3\times3$ mixing matrix by one row and column. In the two-flavor approximation the oscillation probability is given by 
\begin{equation}
    P(\nu_\alpha\to\nu_\beta)=\delta_{\alpha\beta}-4|U_{\alpha 4}|^2(\delta_{\alpha\beta}-|U_{\beta 4}|^2)\sin^2(\Delta m^2 L/4E)
\label{eq:sterile_short_baseline}
\end{equation}
where $\alpha,\beta=e,\mu,\tau$, and $U_{\alpha 4}$ denotes the active components of the new neutrino mass state, $\nu_4$. If the LSND anomaly is indeed produced by an eV-scale sterile neutrino oscillations $\bar\nu_\mu\to\bar\nu_e$, signals must be present in both muon and electron neutrino disappearance searches. 

{\bf Neutrino-electron appearance searches}. The MiniBooNE experiment was designed to search for appearance of electron neutrinos arising at the same $L/E$ as LSND. Recently the MiniBooNE experiment reported a significant excess of neutrino-electron-like events over background compatible with the LSND anomaly in the neutrino and antineutrino modes~\cite{Aguilar-Arevalo:2018gpe}. The MicroBooNE experiment, operating at the same beamline as MiniBooNE, is targeted on confirming if the excess is indeed due to electrons and not mismodeled backgrounds or processes, e.g.~photon emission. Next generation neutrino-electron appearance using pion and kaon decay at rest neutrino fluxes will be performed by the $JSNS^2$ experiment, providing a third check to the LSND claim. The array of detectors that make up the short-baseline neutrino program at Fermilab will also be able to perform very precise measurements of neutrino-electron appearance.

{\bf Neutrino-muon disappearance searches}. MINOS+ has searched for muon-neutrino disappearance in the NuMI baseline and has found no evidence, placing strong constraints on eV-scale sterile neutrinos~\cite{Adamson:2017uda}. Future muon-neutrino disappearance experiments using human-made sources consist of the SBN program, DUNE, and T2HK. IceCube has also searches for muon-neutrino disappearance induced by sterile neutrinos in the TeV-energy range using atmospheric neutrinos. In the latter case the oscillation probability is modified due to the matter potential difference between active and sterile neutrinos; see e.g. Refs.~\cite{Nunokawa:2003ep, Esmaili:2013cja, Petcov:2016iiu} for further details. 
Muon neutrino disappearance using atmospheric neutrinos is also possible in the sub-100 GeV regime and limits on the parameter space of $|U_{\mu 4}|^2$ and $|U_{\tau 4}|^2$ have been placed by Super-Kamiokande~\cite{Wendell:2014dka}, ANTARES~\cite{Albert:2018mnz}, and IceCube-DeepCore~\cite{Aartsen:2014yll}.
Next generation neutrino telescopes such as Km3Net~\cite{Adrian-Martinez:2016fdl} and GVD~\cite{Avrorin:2019dli} will perform similar searches as IceCube high-energy analysis with an improved energy resolution. 
At the same time, next generation atmospheric detectors such as Hyper-K and DUNE will produce an improve characterization of the atmospheric neutrino oscillation.

{\bf Neutrino-electron disappearance searches:} A revisit of the neutrino reactor flux in~\cite{Mention:2011rk}, hinted that there was an overall deficit of electron neutrinos in the global reactor data. This is known as the reactor antineutrino anomaly (RAA). The RAA can also be explained in terms of disappearance induced by sterile neutrinos. Uncertainties in the reactor neutrino flux suggest a reduction in the strength of this anomaly~\cite{RENO:2018pwo,An:2017osx}, but the effects of flux uncertainty presently appear insufficient to explain the full extent of the observed disappearance~\cite{Hayes:2017res,Ashenfelter:2018jrx}. Recent reactor experiments have measured the electron-neutrino flux as a function of distance. Results from DANSS, NEOS, STEREO, and PROSPECT are in tension with the sterile neutrino interpretation of the RAA. On the other hand, these experiments have themselves new anomalous neutrino electron disappearance at the $\sim3\,\sigma$ level.
Large intensity neutrino-electron beams would be needed to further test neutrino electron disappearance. Beta beams required for this purpose have been studied as part of the IsoDAR experiment proposal. 
In addition, there is an observed deficit in the number of neutrinos measured from radioactive sources by GALLEX~\cite{Hampel:1997fc} and SAGE~\cite{Abdurashitov:2009tn} known as the Gallium anomaly~\cite{Giunti:2010zu} that is in a similar parameter space as the RAA.

\begin{figure}
    \centering
    \includegraphics[width=0.3\textwidth]{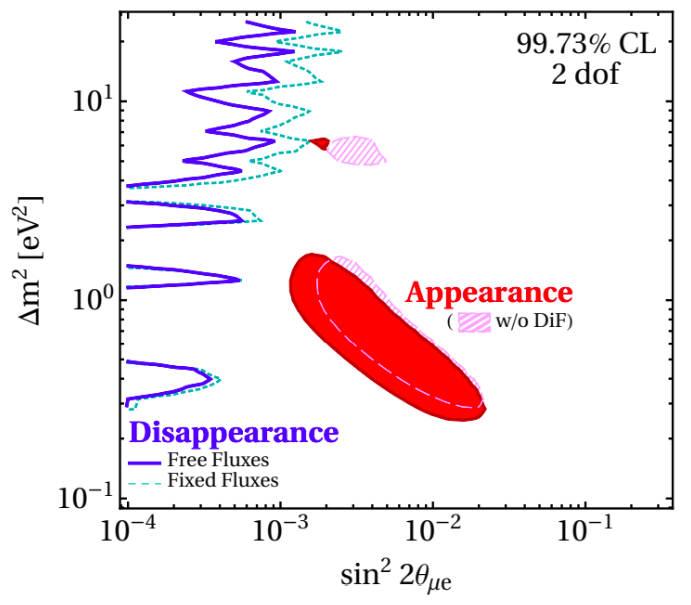}~ 
    \includegraphics[width=0.335\textwidth]{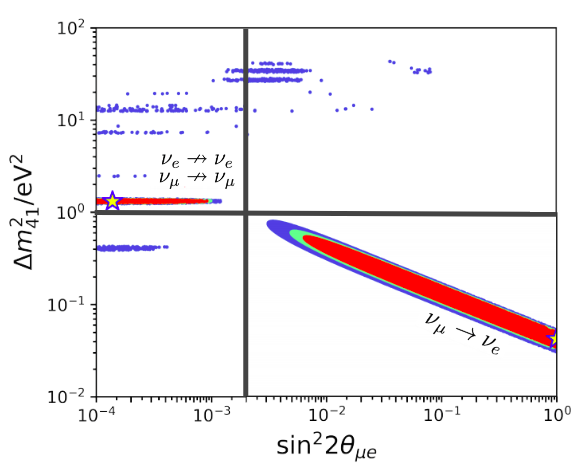}~
    \includegraphics[width=0.3\textwidth]{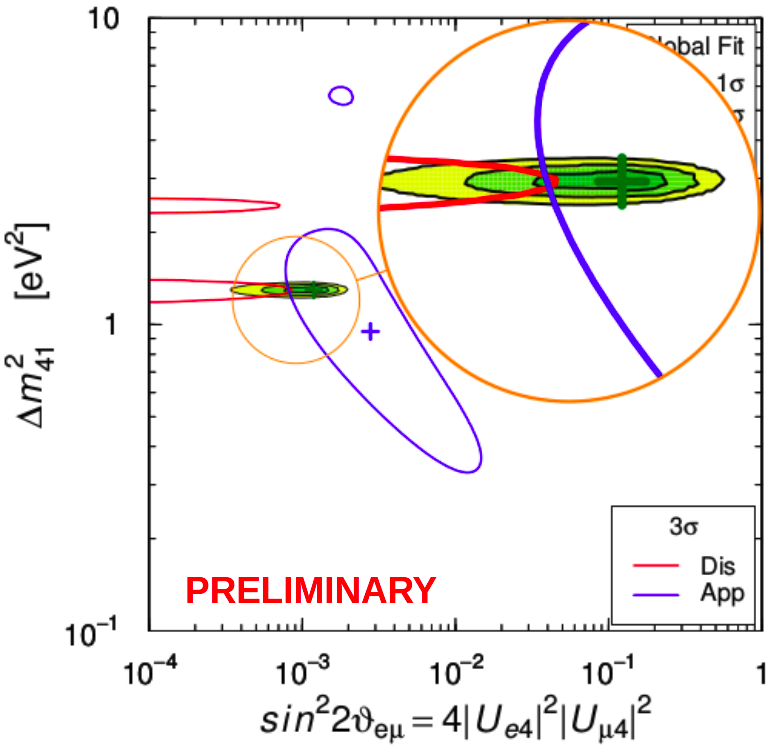}
    \caption{The $\nu_{\mu}\to\nu_e$ appearance amplitude in a 3+1 model is shown in the horizontal axis. 
    Allowed or preferred regions for appearance and disappearance global combinations from three different global-fit groups are shown in left~\cite{Dentler:2018sju}, center~\cite{Diaz:2019fwt}, and right~\cite{ternes_christoph_andreas_2018_2642337,Giunti:2019aiy} plots. See the references for the details.
    }
    \label{fig:app_dis_tension}
\end{figure}

Global combinations of the above channels have been performed by multiple global-fit groups~\cite{Dentler:2018sju,ternes_christoph_andreas_2018_2642337,Giunti:2019aiy,Diaz:2019fwt}.
All groups find that there is significant tension between the appearance and disappearance results as shown in Fig.~\ref{fig:app_dis_tension}, which places a severe burden on the 3+1 model.
Attempts to alleviate this tension, in the context of the 3+1 model, have been studied; {\it e.g.} by adding non-standard interactions~\cite{Liao:2016reh,Liao:2018mbg,Esmaili:2018qzu,Denton:2018dqq} or allowing the sterile neutrino to decay~\cite{PalomaresRuiz:2005vf,Moss:2017pur,Diaz:2019fwt}. 
\subsubsection{Non-standard Neutrino Interactions}
\label{sec:nsi}
Non-standard neutrino interactions (NSIs) provide a general effective field theory (EFT) framework to parameterize new physics in the neutrino sector~\cite{Wolfenstein:1977ue} (see also Refs.~\cite{Guzzo:1991hi, Grossman:1995wx}).
NSIs can be neutral current (NC) or charged current (CC) interactions with matter particles, $e$, $u$, and/or $d$.
Charged-current NSI leads to modification of both the production and detection of neutrinos, but also leads to charged-lepton flavor violation, 
strongly constrained, e.g., from $\mu\to e\gamma$ \cite{TheMEG:2016wtm}.
While NC NSI (with charged leptons) does not modify production, it does affect oscillation and detection.
While NSIs typically refer to vector mediators, scalar mediators~\cite{Ge:2018uhz} or other spin structures~\cite{AristizabalSierra:2018eqm} can also be explored.
Beyond the realm of an EFT approach or a simplified model approach, UV complete NSI models also exist \cite{Davoudiasl:2011sz, Heeck:2011wj,Farzan:2015hkd, Farzan:2016wym,Babu:2017olk,Denton:2018dqq, Wise:2018rnb}.

\begin{figure}
    \centering
    \includegraphics[width=5in,height=2.5in]
    {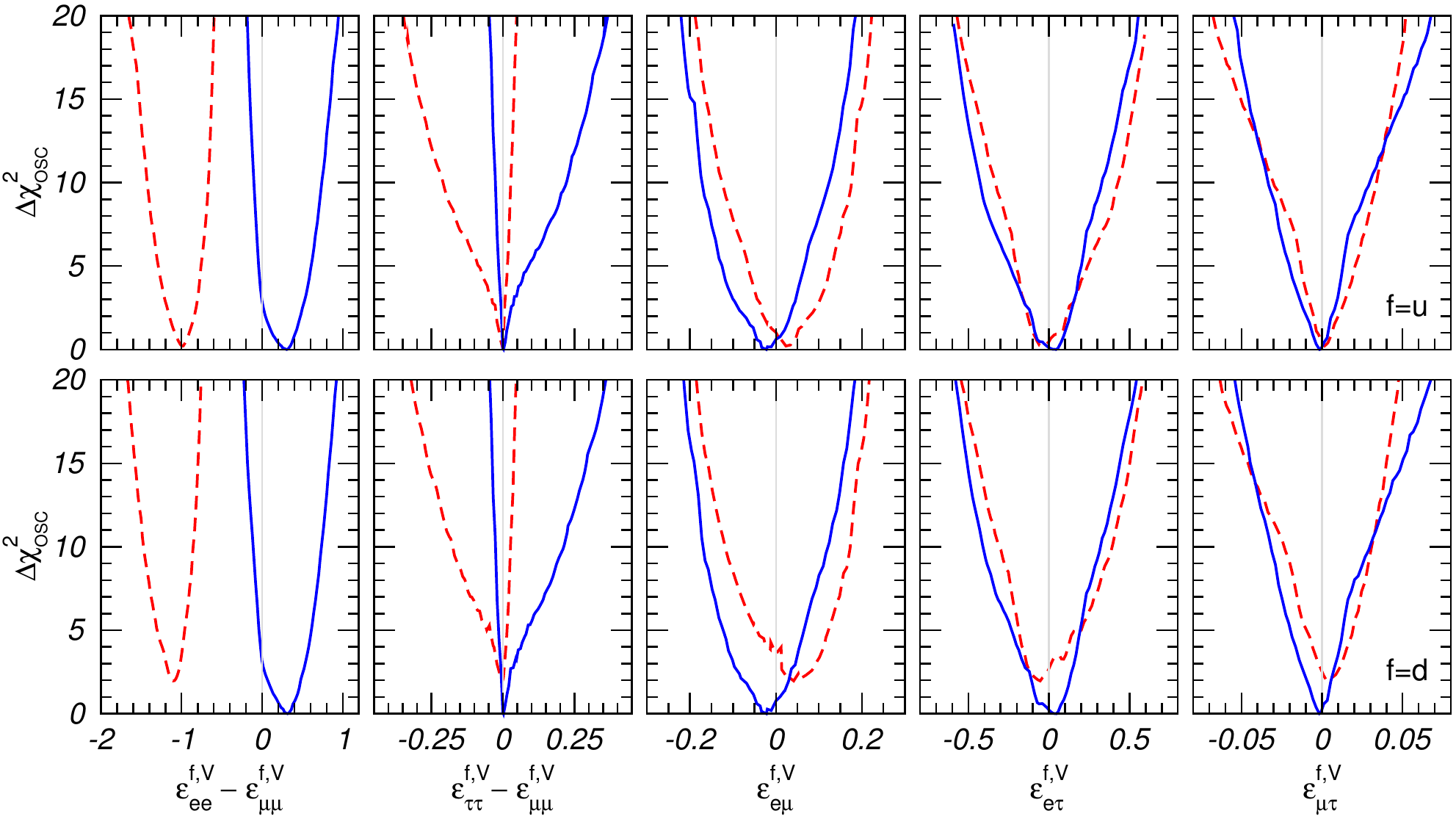}
    \caption{Constraints on NSI terms from oscillation data alone taken one term at a time; the upper (lower) row corresponds to NSI with up (down) quarks.
    Blue: standard solution with $\theta_{12}<45^\circ$. Red: degenerate solution with $\theta_{12}>45^\circ$ and large $|\epsilon_{ee}|\sim1$ NSI \cite{Coloma:2017egw}.
    }
    \label{fig:nsi constraints}
\end{figure}

In the context of propagation, NSIs result in the following modification of the oscillation Hamiltonian,
\begin{equation}
H=\frac1{2E}U_{\rm PMNS}
\begin{pmatrix}
0\\&\Delta m^2_{21}\\&&\Delta m^2_{31}
\end{pmatrix}
U_{\rm PMNS}^\dagger+\sqrt2G_F N_e
\begin{pmatrix}
1+\epsilon_{ee}&\epsilon_{e\mu}&\epsilon_{e\tau}\\
\epsilon_{e\mu}^*&\epsilon_{\mu\mu}&\epsilon_{\mu\tau}\\
\epsilon_{e\tau}^*&\epsilon_{\mu\tau}^*&\epsilon_{\tau\tau}\,,
\end{pmatrix}\,,
\end{equation}
where $G_F$ is the Fermi constant, $N_e$ is the electron density in matter, and the $\epsilon$ terms parameterize the scale of the new interaction in terms of the weak interaction, with respect to the electron-number density. The $\epsilon$ limits can be re-scaled in order to consider interactions with $u$ or $d$.
While some constraints are at the $\sim$ few~\% level, some terms can be about as large as the weak interaction ($|\epsilon|\sim1$), in particular $\epsilon_{ee}$ due to an exact degeneracy in the Hamiltonian \cite{deGouvea:2000pqg,Miranda:2004nb,Coloma:2016gei} that can only be broken by measurements in matter with different neutron densities such as the sun and the Earth.
In addition, $\epsilon_{e\tau}$ is also quite unconstrained due to comparatively few appearance measurements, see Fig.~\ref{fig:nsi constraints}. Next generation oscillation experiments are expected to improve the current constraints, although some partial degeneracies will still exist \cite{GonzalezGarcia:2001mp,Friedland:2012tq, Girardi:2014kca, Masud:2015xva,Coloma:2015kiu,deGouvea:2015ndi,Liao:2016hsa,deGouvea:2016pom,Ge:2016dlx,Agarwalla:2016fkh,Blennow:2016etl,Fukasawa:2016lew,Deepthi:2017gxg,Hyde:2018tqt}. The leading constraints on vector NSI parameter $\epsilon_{e\mu}$ come from IceCube, IceCube-DeepCore, and Super-Kamiokande.
Improved constraints on vector NSI ought to be possible in next-generation atmospheric neutrino experiments such as Km3Net, GVD, and IceCube-Upgrade, as well as at DUNE~\cite{Coloma:2015kiu,Masud:2017bcf} and HK~\cite{Liao:2016orc,Kelly:2017kch}. Scalar NSI effects 
can be studied with e.g.~JUNO and GeV atmospheric neutrinos.

Non-standard interactions can also be probed at scattering experiments for mediator masses $\gtrsim$ the energy scale of the experiments such as NuTeV~\cite{Zeller:2001hh,Ball:2009mk,Bentz:2009yy,Coloma:2017egw} for mediator mass $\gtrsim 10$ GeV and COHERENT~\cite{Akimov:2017ade} probing the coherent elastic neutrino nucleus scattering (CE$\nu$NS) channel \cite{Freedman:1973yd} for mediator mass even down to $\sim10$ MeV \cite{Liao:2017uzy,Denton:2018xmq}.
Future CE$\nu$NS measurements 
can push this limit down by about one order of magnitude by leveraging even lower threshold detectors and a lower energy flux of neutrinos \cite{Wong:2005vg,Aune:2005is,Brudanin:2014iya,Lindner:2016wff,Aguilar-Arevalo:2016qen,Agnolet:2016zir,Billard:2016giu}.
\subsubsection{Neutrino Tridents}

\begin{figure}
    \centering
    \includegraphics[width=0.5\textwidth]
    {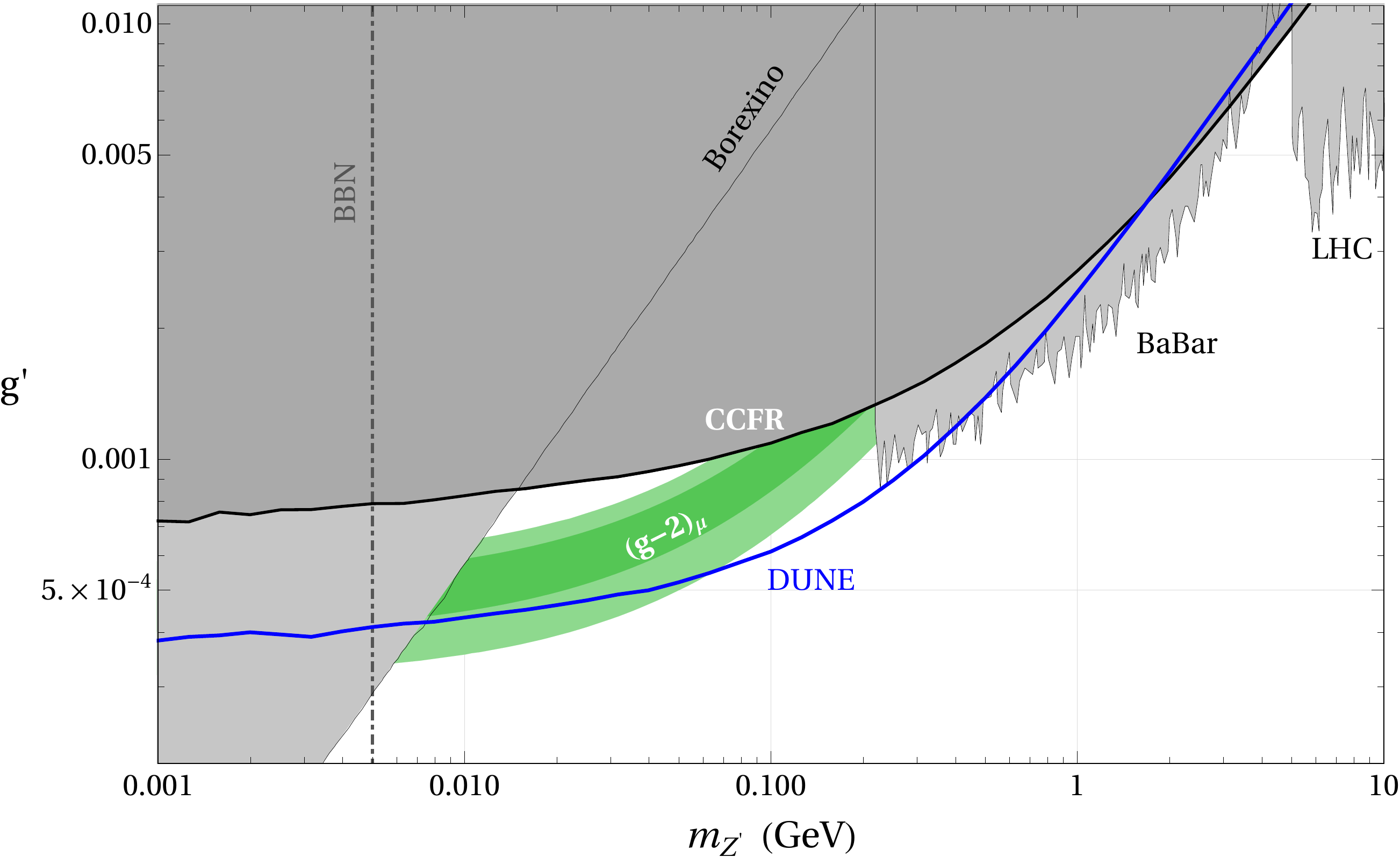}
    \caption{Projected DUNE sensitivity in the $L_\mu - L_\tau$ parameter space. The green region shows where the $(g-2)_\mu$ anomaly can be explained at the $2\sigma$ level. See Ref.~\cite{Altmannshofer:2019zhy} for the details.
    }
    \label{fig:tridents1}
\end{figure}

Neutrino trident production is a rare weak process in which a neutrino, scattering off the Coulomb field of a heavy nucleus, generates a pair of charged leptons~\cite{Czyz:1964zz,Lovseth:1971vv,Fujikawa:1971nx,Koike:1971tu,Koike:1971vg,Brown:1973ih,Belusevic:1987cw,Ballett:2018uuc}. The typical final state of a neutrino trident interaction contains two leptons of opposite charge. Measurements of dimuon tridents ($\nu_\mu \to \nu_\mu \mu^+\mu^-$) have been carried out at the CHARM-II~\cite{Geiregat:1990gz} and CCFR~\cite{Mishra:1991bv} experiments. Later, a dimuon measurement at NuTeV~\cite{Adams:1999mn} was shown to be consistent with no trident events at the 1$\sigma$ level. These results are consistent with Standard Model (SM)  predictions, but leave ample room for improvement as well as for potential contributions from new physics. Beyond measuring dielectron and mixed flavor channels for the first time, one can envisage probing the SM weak interactions through more precise measurements of dimuon tridents, where interference effects between the CC and NC are significant. In fact, it was recently shown that a class of $Z'$ models can modify the dimuon trident cross sections. These models introduce a $Z'$ boson by gauging an anomaly-free global symmetry of the SM, with a particular interesting case realized by gauging $L_\mu - L_\tau$~\cite{He:1990pn,He:1991qd}. Such a $Z'$ is not very tightly constrained and could address the observed discrepancy between the SM prediction and measurements of the anomalous magnetic moment of the muon, $(g-2)_\mu$~\cite{Baek:2001kca,Harigaya:2013twa}.
The near detectors of future neutrino experiments offer an excellent environment to generate a sizeable number of trident events, offering excellent prospects to both improve the above measurements and to look for an excess of events above the SM prediction, which would be an indication of New Physics. Figure~\ref{fig:tridents1} shows the expected sensitivity of DUNE to the $Z'$ model described above through a dimuon trident measurement. Finally, such $Z^\prime$ models can also impact neutrino-electron scattering measurements, providing an additional handle in the search for new physics at the DUNE ND~\cite{Ballett:2019xoj}. 

Trident measurements also serve as a probe of several other dark sectors at neutrino experiments. Charged lepton pairs may arise from the decay-in-flight of dark photons, dark scalars or sterile neutrinos, from dark bremsstrahlung in neutrino interactions~\cite{deGouvea:2018cfv} or from upscattering into highly unstable dark states. The latter scenario, for instance, typically arises in dark neutrino sectors where SM neutrinos mix with heavier SM singlet fermions (dark neutrinos) to induce BSM interactions.
The upscattered heavy states decay back to a pair of charged-leptons giving trident signatures of type $\nu A\to N A\to \nu\, \ell^+\ell^- A$.
These scenarios may explain the smallness of neutrino masses~\cite{Bertuzzo:2018ftf,Ballett:2019cqp} and possibly the MiniBooNE low energy excess of electron-like events~\cite{Bertuzzo:2018itn,Ballett:2018ynz,Arguelles:2018mtc,Ballett:2019pyw,Coloma:2019qqj}.

\subsubsection{Large Extra-Dimensions}
%
The model of Large Extra-Dimensions (LED) in Ref.~\cite{Davoudiasl:2002fq}, which is based on seminal works that also focused on the generation of the neutrino mass,  
implies right-handed neutrinos propagating in the extra dimensions that couple 
to the active neutrinos
which thus acquire a Dirac Mass in this framework. After compactification of the large extra dimension, an infinite number of sterile neutrinos appear from the Kaluza Klein modes $n$ in the bulk.
The sterile-active mixing and the new oscillation frequencies modify the three flavor active neutrino oscillations and hence the absolute mass scale $m_0$ and the compactification radius $R$ are constrained~\cite{Davoudiasl:2002fq,Machado:2011jt, Machado:2011kt,Esmaili:2014esa,DiIura:2014csa, Berryman:2016szd,Stenico:2018jpl}.  
For $n=0$, which corresponds to the more active case, and for $m^D\,R \ll 1$, where $m^D$ denotes the Dirac neutrino mass, the standard three neutrino oscillations should be recovered. 
The more sterile case, i.e., $n\gg1$, oscillations will appear smeared at the far detector since large $n$ implies large oscillation phases and also the active-sterile mixing will be suppressed.

\begin{figure}[ht]
\centerline{
\includegraphics[width=0.5\textwidth]
{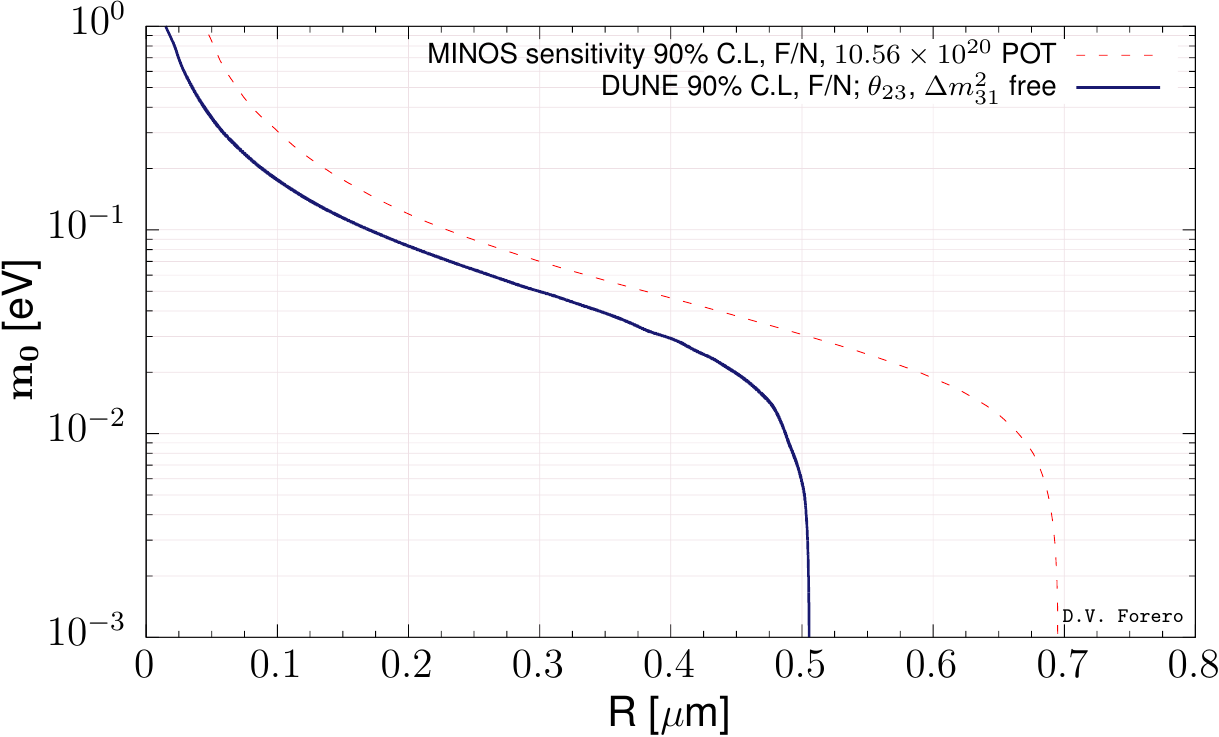}
}
\caption{\label{fig:ledsensitivity} DUNE sensitivity to the free parameters of the LED model in Ref.~\cite{Davoudiasl:2002fq} through its impact on the neutrino oscillations expected in DUNE. See text for details of the analysis.}
\end{figure}

Figure~\ref{fig:ledsensitivity} shows the DUNE and MINOS~\cite{Adamson:2016yvy} sensitivities to LED at $90\%$ of C.L. for 2 d.o.f given by the full and dashed lines, respectively. 
In the case of DUNE, an exposure of $300\,\text{kt}\,\text{MW}\, \text{year}$  was assumed and spectral information from the four oscillation channels, (anti)neutrino appearance and disappearance, were included in the analysis. 
In the analysis, the `true' event energy spectrum corresponds to the standard three neutrino case (which is recovered in the limit $R\to0$) and the LED model parameters are then fitted. In the fit, the solar parameters were kept fixed, and also the reactor mixing angle, while the atmospheric parameters were considered as free. For an earlier study of the DUNE sensitivity to LED model free parameters see Ref.~\cite{Berryman:2016szd}. 
In general, DUNE improves over the MINOS sensitivity for all values of $m_0$ and this is more noticeable for the lowest $m_0$ values where a conservative sensitivity limit to $R$ is obtained. 

\subsubsection{High Energy Cross Sections}
An important test of the SM and, as such, a probe of BSM neutrino physics, comes in the form of measuring the neutrino-nucleon cross section at extreme energies.
The highest energy cross sections from accelerators are at $E_\nu\sim350$ GeV \cite{Tanabashi:2018oca}.
The only higher energy cross section measurement comes from IceCube at energies of $E\sim100$ TeV \cite{Aartsen:2017kpd,Bustamante:2017xuy}.
Upgrades of IceCube and KM3NeT can improve the measurement for 10 TeV $\lesssim$ E $\lesssim$ 1 PeV and future radio experiments such RNO, POEMMA, and GRAND \cite{Alvarez-Muniz:2018bhp,Barwick:2016mxm,Olinto:2017xbi,Allison:2014kha} could push this to the EeV scale.
Finally, LHC neutrino experiments such as FASER \cite{Feng:2017uoz,Abreu:2019yak} could measure the neutrino-nucleon cross section at $E\sim1$ TeV filling in the energy gap between accelerator measurements and neutrino telescope measurements.

Such a measurement provides not only a check of the neutrino cross section at energies not accessible anywhere else, but it can also probe new physics models such as extra dimensions \cite{Connolly:2011vc}, leptoquarks \cite{Romero:2009vu}, sphalerons, and others \cite{Klein:2019nbu}.
These kinds of measurements also provide an opportunity to tomographically measure the Earth and probe its weak charge \cite{Donini:2018tsg}, something that could also be significantly improved at upcoming experiments.

\subsubsection{BSM Physics with Tau Neutrinos}
%
To date, the only identified charged-current $\nu_\tau$ scattering events have been measured by the DONUT~\cite{Kodama:2000mp} and OPERA~\cite{Agafonova:2018auq,Agafonova:2019npf}. Other experiments have statistically identified a tau event sample, such as Super-Kamiokande~\cite{Li:2017dbe} and IceCube~\cite{Aartsen:2019tjl}. In the next generation of neutrino experiments, DUNE and HK are expected to collect a large sample of these events.
In seven years of data collection, $\mathcal{O}(10^3)$ $\nu_\tau$ charged-current events are expected to occur in the DUNE far detector.
Using the methods developed in Ref.~\cite{Conrad:2010mh}, the DUNE collaboration expects to identify roughly 30\% of hadronically-decaying $\tau$ leptons and only having $\sim 0.5\%$ of neutral current scattering events as background contamination.

\begin{figure}
\centerline{
\includegraphics[width=0.9\linewidth]{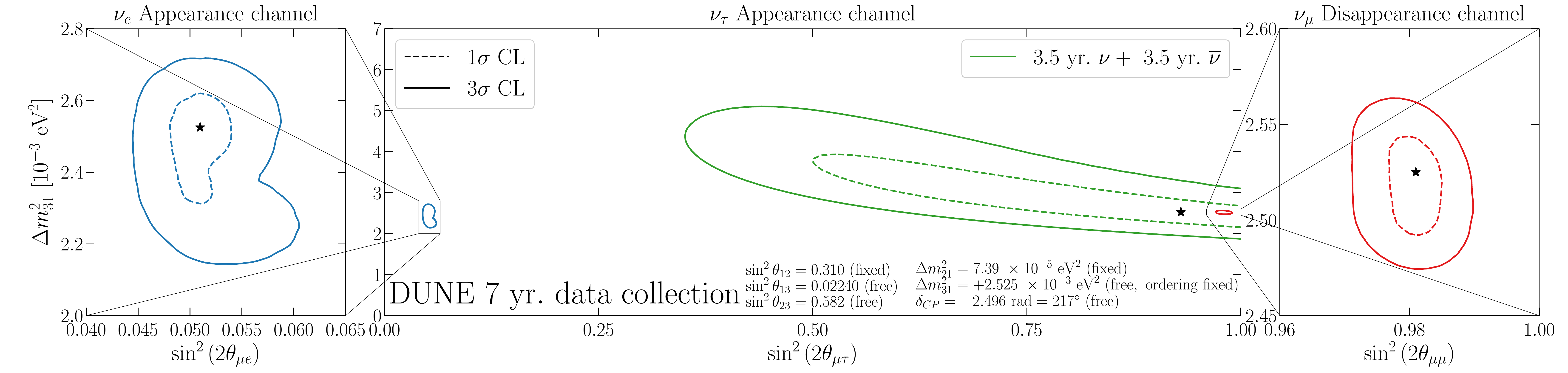}
}
\caption{\label{fig:EffectiveAngles_Tau} Expected measurement potential of seven years of data collection at DUNE, assuming separate analyses of $\nu_e$ appearance (left), $\nu_\tau$ appearance (center), or $\nu_\mu$ disappearance (right). Each panel displays $1\sigma$ (dashed) and $3\sigma$ (solid) CL regions of the measurement of the mass splitting $\Delta m_{31}^2$ and the effective mixing angle $\sin^2(2\theta_{\mu \beta})$ (see text for detail). The solar neutrino parameters have been fixed, but all other parameters have been marginalized in the fit.}
\end{figure}

\begin{figure}[ht]
\centerline{
\includegraphics[width=0.4\textwidth]{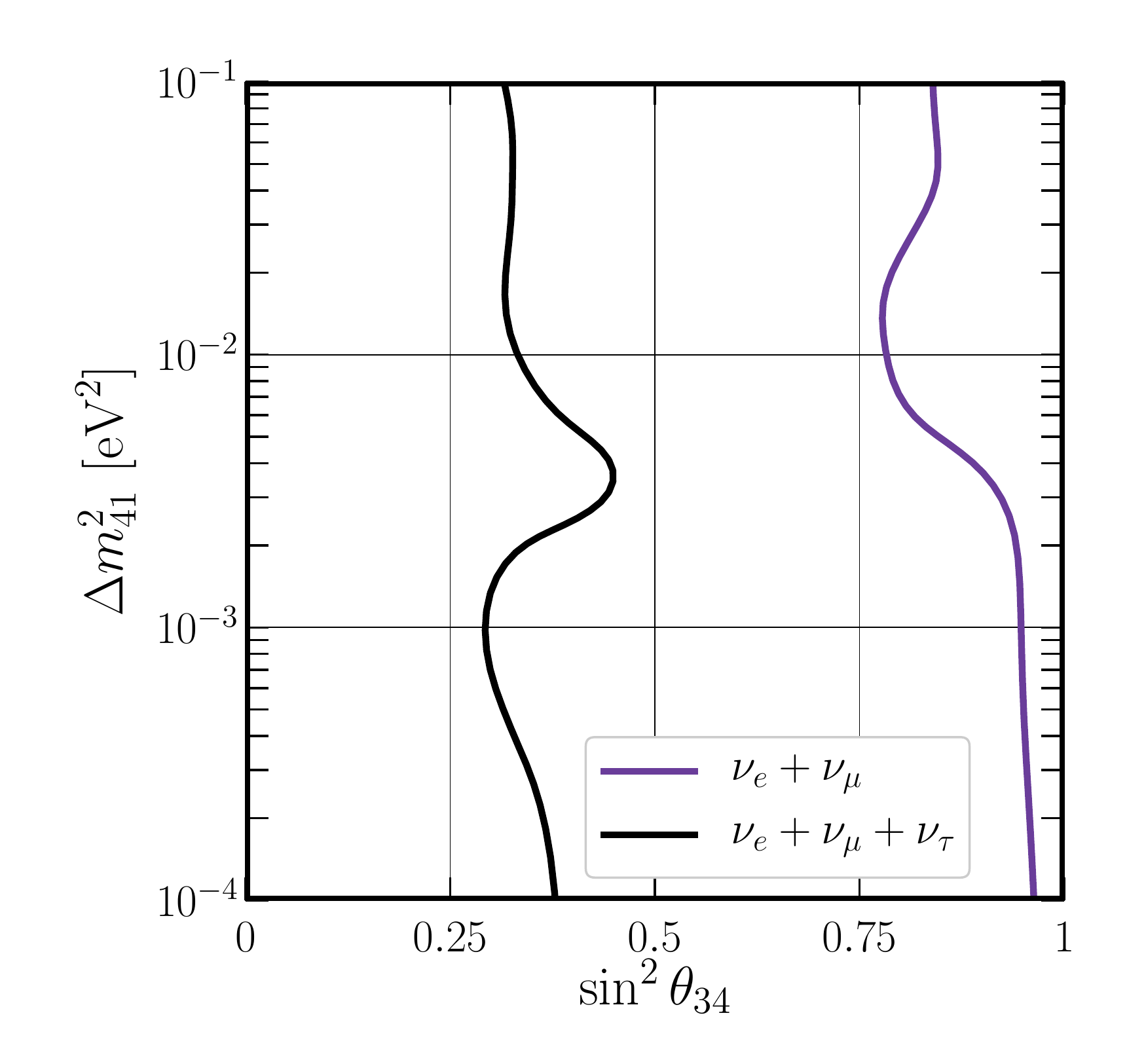}
}
\caption{\label{fig:SterileTau} Expected DUNE sensitivity (7-year data) to the sterile neutrino mixing parameter $\sin^2\theta_{34}$ and new mass-squared splitting $\Delta m_{41}^2$ assuming data collection using only the $\nu_e$ appearance and $\nu_\mu$ disappearance channels (purple) or a combined analysis including the $\nu_\tau$ appearance channel (black). 
See Refs.~\cite{deGouvea:2019ozk,Berryman:2015nua} for more detail.}
\end{figure}

Reference~\cite{deGouvea:2019ozk} explored the capabilities of performing physics measurements with this $\nu_\tau$ sample, both in the three-massive-neutrinos paradigm and in the context of BSM 
scenarios such as a non-unitary lepton mixing matrix, the existence of neutrino nonstandard interactions, and the existence of a light sterile neutrino.
While not as powerful as the $\nu_e$ appearance and $\nu_\mu$ disappearance channels at DUNE, this sample allows for complementary measurements that operate as a consistency check. Figure~\ref{fig:EffectiveAngles_Tau} displays the capability of each channel individually measuring the effective mixing angle for neutrino oscillations and the atmospheric mass splitting at DUNE. This will be the first such measurement using $\nu_\tau$ events.
In Fig.~\ref{fig:EffectiveAngles_Tau}, the mixing angle $\sin^2(2\theta_{\mu\beta})$ is defined as $4|U_{\mu 3}|^2 |U_{\beta 3}|^2$ (for $\beta \neq \mu$) or $4|U_{\mu 3}|^2 (1 - |U_{\mu 3}|^2)$ (for $\beta = \mu$).
Figure~\ref{fig:SterileTau} displays the expected DUNE sensitivity to the sterile neutrino mixing parameter.

\subsubsection{Non-Unitarity}
%

A popular mechanism to generate neutrino masses is the so called type-I see-saw model where extra singlet fermions are introduced to the SM. 
In the low-scale version of this models the energy scale of the heavy leptons can be set around the TeV.  
In this low-scale see-saw models~\cite{Mohapatra:1986bd,Akhmedov:1995ip,Malinsky:2005bi} sizable deviation of the effective lepton mixing matrix unitarity can be at the percent level~\cite{Forero:2011pc} and therefore with the potential to impact neutrino oscillations. 
While mentioned heavy leptons will be decoupled from neutrino oscillations, its mixing will be given by a non-unitary lepton mixing matrix, which can be conveniently parametrized as a lower triangular matrix~\cite{Escrihuela:2015wra} (of dimension three), correcting the usually assumed unitary neutrino mixing matrix~\cite{Tanabashi:2018oca}. 
Direct searches of such heavy leptons could be a test of non-unitarity and they can be performed at next-generation experiments.

The neutrino oscillations with a non-unitarity neutrino mixing matrix have recently received attention, see for instance 
Refs.~\cite{Escrihuela:2015wra,Li:2015oal,Fong:2016yyh,Pas:2016qbg,Escrihuela:2016ube,Blennow:2016jkn}. 
Electroweak precision tests and charged lepton flavor violation (cLFV) constraints reduce the margin for unitarity deviation in model dependent ways~\cite{Langacker:1988ur,Nardi:1994iv,Antusch:2006vwa,deGouvea:2015euy,Antusch:2014woa,Fernandez-Martinez:2016lgt}.
Adopting a more model independent approach, and separating constraints from neutrino oscillations (see for instance Ref.~\cite{Parke:2015goa}) to the ones involving also charged leptons (electroweak and cLFV), the impact of having a non-unitary lepton mixing matrix in the CP sensitivity in the DUNE experiment has been considered in Ref.~\cite{Escrihuela:2016ube}. 
Also, a very small decrease of the CP sensitivity results from the freedom in the unknown new phases (additional to the Dirac CP phase) encoded in the lower triangular matrix accounting for non-unitarity.

\subsubsection{Lorentz Violation}
%
Violation of Lorentz and CPT symmetry may be present in  extensions of the SM like string theory~\cite{Kostelecky:1988zi}.
In scenarios where Lorentz/CPT symmetries are broken dynamically, a non-trivial space-time dependence of the vacuum of the theory, such as a preferred direction, leads to an apparent violation of these symmetries.
In neutrino experiments, Lorentz/CPT violation may arise as modifications of oscillation probabilities, like time or direction dependent effects, neutrino-antineutrino mixing, or energy dependent effects on mass splittings~\cite{Kostelecky:2003cr, Kostelecky:2004hg, Katori:2006mz, Diaz:2011ia}. The effect of Lorentz/CPT violation can be organized in terms of effective operators~\cite{Kostelecky:2011gq} 
\begin{equation}
H\sim\frac{m^2}{2E}+\accentset{\circ}{a}^{(3)}-E\cdot \accentset{\circ}{c}^{(4)} + E^2\cdot\accentset{\circ}{a}^{(5)} -E^3\cdot\accentset{\circ}{c}^{(4)}\cdots .
\end{equation}
where the $\accentset{\circ}{a}^{(n)}$ is the strength of Lorentz violation induced by a CPT-odd interaction with using a n-dimension operator to relate the LV-field and the neutrino field, while the $\accentset{\circ}{c}^{(n)}$ represent the same thing for CPT-even interactions. The lowest-order interaction has the similar phenomenology as an NSI in constant density~\cite{Diaz:2015dxa,Barenboim:2018lpo}, while higher-order operators effects grow as the neutrino energy increases. In fact, current measurements of the difference between neutrinos and antineutrino oscillation probability parameters already provides some of the strongest test of CPT invariance~\cite{Barenboim:2017ewj}.

The oscillation probability scales like $P(\accentset{\circ}{a}^{(n)} L E^{n-3})$ and hence  higher-dimensional operators are better constrained by the measurements of high-energy atmospheric neutrino~\cite{Aartsen:2017ibm} and astrophysical neutrino flavor ratio~\cite{Arguelles:2015dca}.
Lorentz violation can also induce time variability of the neutrino oscillation~\cite{Abbasi:2010kx} where lower-dimensional operators can be constrained from oscillation experiments~\cite{Adamson:2008aa, Adamson:2010rn, Abbasi:2010kx, Mitsuka:2011ty,  AguilarArevalo:2011yi, Abe:2012gw, Albert:2016hbz,  Aartsen:2017ibm, Abe:2017eot, Kumar:2017sdq, Adey:2018qsd, Barenboim:2018ctx}.

Lorentz violation effects can produce kinematic effects, e.g. neutrino bremsstrahlung to an 
$e^+ e^-$ pair is allowed in the case of superluminal neutrinos, 
searched by time-of-flight measurements~\cite{Longo:1987ub, Adamson:2015ayc} and vacuum cherenkov emissions~\cite{ICARUS:2011aa,Cohen:2011hx}. 
Accordingly, high-energy astrophysical neutrinos rapidly lose energy and show a characteristic cuf-off in the energy spectrum~\cite{Stecker:2017gdy}. 
Lorentz violation can be also probed in neutrinoless double beta decay experiments~\cite{Albert:2016hbz} as well as those with xenon gas TPCs.

\subsubsection{Heavy Neutral Leptons}
%
Heavy Neutral Leptons (HNLs, also sometimes called heavy sterile neutrinos) with masses below the electro-weak scale are a prime target for current and next generation neutrino experiments. Such particles can be directly tied to the generation of neutrino masses and may also play an important role in the dark matter and the matter-antimatter asymmetry puzzles~\cite{Akhmedov:1998qx,Asaka:2005pn}. 

HNLs lead to a rich set of signatures that can be searched for in a variety of experiments~\cite{Atre:2009rg, Drewes:2013gca}. Accelerator-based neutrino experiments with near detectors provide one promising venue for HNL searches 
because HNLs, produced through meson rare decays, can travel to the near detector and decay to SM particles, e.g., a lepton and a meson or three leptons.
It should be pointed out that if these HNLs have additional interactions, e.g.\ due to new dark gauge sector,
the phenomenology can be significantly modified~\cite{Ballett:2019cqp} and dedicated searches should be carried out.

A search for HNL decays in the T2K near detector ND280 has been negative and has yielded constraints~\cite{Abe:2019kgx}. Furthermore, searches for HNLs have been proposed at IceCube~\cite{Coloma:2017ppo}, at the Fermilab SBN experiments~\cite{Ballett:2016opr}, and a search is currently underway within MicroBooNE.
Beyond conventional neutrino experiments, proposed experiments such as SHiP~\cite{Anelli:2015pba} and MATHUSLA \cite{Curtin:2018mvb}, as well as the recently approved FASER \cite{Feng:2017uoz} are dedicated long lived particle detectors that can probe interesting regions in the HNL mass - mixing angle parameter space~\cite{Beacham:2019nyx,Alimena:2019zri}.
DUNE ND, thanks to the high intensity of the beam and the large near detector complex, %
can achieve a competitive sensitivity to HNL with masses below GeV as other dedicated experiments~\cite{Ballett:2019bgd}.
If an HNL is discovered, the study of its decays could tell us whether all neutrinos are Majorana particles or Dirac particles~\cite{Balantekin:2018ukw}.

\begin{figure}
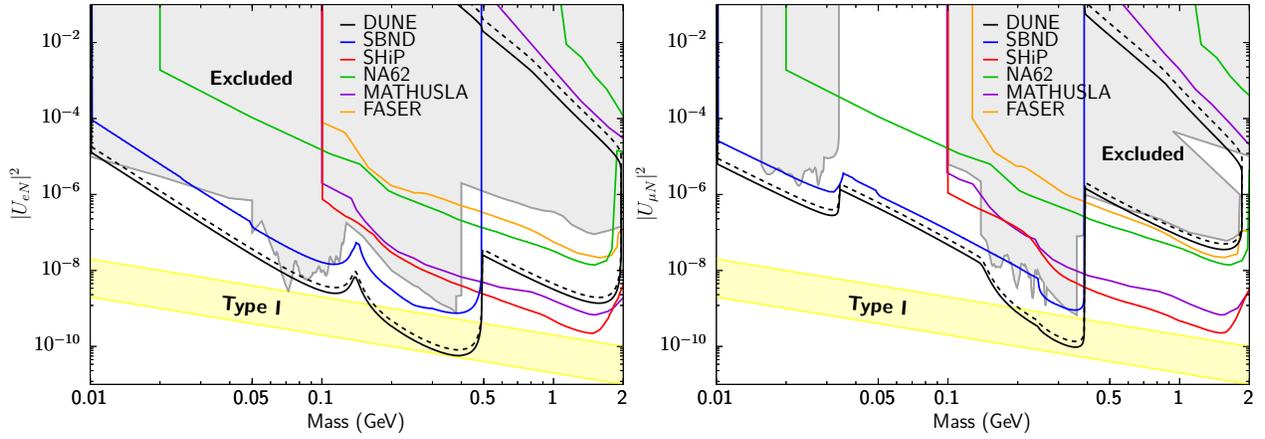

    \centering
    {\resizebox{0.45\linewidth}{!}{\input{graphics/DUNE_HNL_E.tex}}}
    {\resizebox{0.45\linewidth}{!}{\input{graphics/DUNE_HNL_M.tex}}}
   \caption{The 90\,\% C.L. sensitivity regions for dominant mixings %
                $|U_{e N}|^2$ (left) and $|U_{\mu N}|^2$ (right)
                are presented~~\cite{Ballett:2019bgd}
                }%
    \label{fig:DUNEND-to-HNL} 
\end{figure}

The experimental signature for these decays in a beam dump experiment is a decay-in-flight event with a decay vertex,
different from a typical neutrino--nucleon scattering, and a rather forward direction with respect to the beam.
The main background to this search comes from SM neutrino--nucleon scattering events %
in which the hadronic activity at the vertex is below threshold.
Charged current quasi-elastic events with pion emission from resonance are background to semi-leptonic decay channels, %
whereas mis-identification of long pion tracks into muon can constitute a background to three-body leptonic decays, %
as the neutrino in the decay products is not reconstructed.
Figure \ref{fig:DUNEND-to-HNL} from Ref.~\cite{Ballett:2019bgd} shows the physics reach of DUNE ND in its current configuration, 
compared to that of SBN, SHiP, NA62, MATHUSLA, and FASER.

%
\subsubsection{Neutrino Decays}
Various new physics models predict neutrino decay with lifetimes probable in various terrestrial and astrophysical experiments. 
Constraints are in terms of $\tau/m$, the lifetime over the mass of the neutrino, and SN1987A places extremely strong constraints on $\bar\nu_e$ decay at $\tau/m\gtrsim10^5$ s/eV~\cite{Hirata:1987hu}.
Constraints from IceCube are at $\tau/m\gtrsim10$ s/eV for all flavors~\cite{Pagliaroli:2015rca,Shoemaker:2015qul,Bustamante:2016ciw}, although a $>3$ $\sigma$ tension points towards neutrino decay with $\tau/m\sim10^2$ s/eV for $\nu_2$ and $\nu_3$ alone as a possible solution~\cite{Denton:2018aml}.
This model also predicts a deficit of $\nu_\tau$'s that could be tested with additional IceCube data.
Solar neutrinos constrain the lifetime of $\nu_2$'s to $\tau/m\gtrsim10^{-3}$ s/eV~\cite{Berryman:2014qha} and long-baseline and atmospheric experiments constrain the lifetime of $\nu_3$ to $\tau/m\gtrsim10^{-9.5}$ s/eV~\cite{GonzalezGarcia:2008ru}.
Finally, the strongest constraints available come from the early universe at $\tau/m\gtrsim10^{11}$ s/eV~\cite{Hannestad:2005ex}.

Numerous models for neutrino decay exist, and most include a very light ($<1$ eV) or massless majoron~\cite{Gelmini:1980re}, which can be related to neutrino mass generation.
Neutrino decay can be visible or invisible~\cite{Lindner:2001fx,Gago:2017zzy,Coloma:2017zpg,Ascencio-Sosa:2018lbk} where invisible means that either the decay products are sterile 
neutrinos, or 
have sufficiently low energy avoiding detection.
Visible neutrino decay involves regeneration of lower energy neutrinos and provides additional detection signatures. 
Other models involving neutrino decay only in the sterile sector \cite{PalomaresRuiz:2005vf} could provide a possible explanation for the light sterile neutrino anomalies.
\subsubsection{Ultra-light Dark Matter}
Ultra-light scalar fields, with masses much below the eV scale, 
arise as key ingredients in several well-motivated extensions of the standard model. 
If such scalar field couples to neutrinos, several novel phenomenological signatures could be observed in neutrino experiments~\cite{Berlin:2016woy, Capozzi:2017auw,Zhao:2017wmo, Krnjaic:2017zlz, Brdar:2017kbt, Liao:2018byh, Capozzi:2018bps, Farzan:2018pnk}.
The occupation value of the scalar field $\phi$, that is, its vacuum expectation value, may modulate in time coherently with a frequency given by the mass of the field $\omega=m_\phi$.
If the scalar field is sufficiently light, $m_\phi\ll \mathcal{\rm days}$, the neutrino mass matrix would modulate in time, leading to a time-modulation of neutrino oscillation probabilities.
The modulation can happen with mixing angles, mass splittings or both, leading to distinctive temporal signatures.
When the period $1/\omega$ is too short to be seen as a time-modulation but still longer than the neutrino time-of-flight, every neutrino produced will be subject to a different, instantaneous mass matrix. 
If there are changes to the mass splittings, this effect will lead to Distorted Neutrino Oscillations (DiNOs), an averaging of oscillation probabilities.
When the period $1/\omega$ is of the order of the neutrino time-of-flight, a non-trivial matter effect takes place.
Other phenomenological impacts of this BSM may also be observed in high energy astrophysical neutrino flavor composition in neutrino telescopes.
\subsubsection{Resonant $\nu_\mu\to\nu_e$ Oscillations.}

By invoking interactions with background particles via light intermediates, matter effects with non-trivial energy dependencies can be generated~\cite{Asaadi:2017bhx}.  
Such models resolve tensions between appearance and disappearance experiments that arise when their oscillations are analyzed universal function of $L/E$.  A light scalar boson coupled only to neutrinos, for example, can introduce forward scattering from the cosmic neutrino background.
For scalar masses in the natural regime for Higgs-like Dirac neutrino mass generation, this produces resonant oscillations in a comparable energy range to the MiniBooNE low energy excess, 
fits MiniBooNE data  slightly better than the 3+1 sterile neutrino model, and significantly alleviates tensions with world data.  
On the other hand, %
justification of a neutrinophilic light Higgs boson is theoretically challenging, and the required over-density of the cosmic neutrino background, 
while not ruled out, is not especially well motivated.  Considered as a general property of models with light scalars, however, energy dependent matter effects with resonances may be considered as compelling phenomenological approaches to resolving short baseline tensions via new physics scenarios, explorable at next generation experiments.

\subsection{Dark Matter and Dark Sector Searches}

\subsubsection{Dark Matter Indirect Detection in Neutrino Experiments}
\label{sec:dmnu}
%
{\bf Dark matter annihilation at the galactic center:} The goal of DM indirect detection is to find the products of its annihilation or decay such as photons, $\nu$, $e^\pm$, and $p^\pm$
, coming from highly concentrated region such as the Galactic Center.
A neutrino travels along an almost straight line from where it is created.
Currently, various large volume experiments are already in operation or planed including DUNE~\cite{Abi:2018rgm, Abi:2018alz}, IceCube~\cite{Aartsen:2016nxy}, and SK~\cite{Fukuda:2002uc}/HK~\cite{Abe:2016ero, Abe:2018uyc}.
Due to their improved sensitivities and large volumes, DUNE will extract more information on DM properties in the near future.
For a given DM model, one can easily calculate the expected neutrino flux from DM annihilation.
Then, the search strategy is to find an excess of neutrinos from, e.g., the Galactic Center direction compared to the expected atmospheric neutrino background. 
No excess of neutrinos has been observed yet and hence provides 
upper limits on DM annihilation cross sections depending on annihilation channels~\cite{Frankiewicz:2015zma, Aartsen:2017ulx, Aartsen:2017mnf}.

{\bf Dark matter annihilation at the Sun:} Dark matter can be accumulated inside the Sun through DM-nuclei and/or DM-DM scatterings~\cite{Gould:1987ir, Damour:1998vg, Chen:2014oaa}, and the captured DM may annihilate into SM particles in the Sun.
Thus, the Sun can be a good point-like source of neutrino flux from dark matter annihilation with a relativiely short distance from the Earth compared to the Galactic Center.
Searches for an excess of neutrinos from the direction of the Sun over the atmospheric neutrino background also have been conducted. 
Upper limits on spin-independent and spin-dependent DM-nucleon scattering cross sections are now available~\cite{Choi:2015ara, Aartsen:2016zhm, Aartsen:2017mnf} since the dark matter capture processes rely on DM-nucleus scattering.
Interestingly, neutrino detectors provide more stringent limits on the spin-dependent DM-proton scattering cross section than those from DM direct detection experiments depending on annihilation channels and dark matter masses. 
\subsubsection{Light Dark Matter and Dark Sectors}
%
While there are in general many interesting and viable scenarios for DM, the paradigm of a light dark sector has risen to prominence in recent years.  
In this framework, dark matter is part of a hidden sector of neutral particles that couples weakly to the SM through a portal interaction involving a light mediator. 
A compelling aspect of this framework is the possibility of a thermal relic origin of dark matter for mass scales as light as ${\cal O}$(MeV)~\cite{Boehm:2003hm}. 
Dark sector models predict a host of novel phenomena,  requiring new experimental approaches beyond those used to search for WIMPs. 
These include new low mass direct detection technologies, sensitive searches for rare decays of SM particles, and fixed target/beam dump experiments; for a nice overview we refer the reader to~\cite{Battaglieri:2017aum}.
Future neutrino experiments will have an important role to play in exploring the space of dark sector theories. 

\begin{figure}[t]
\begin{center}
\includegraphics[width=0.55\textwidth,height=3.5in]{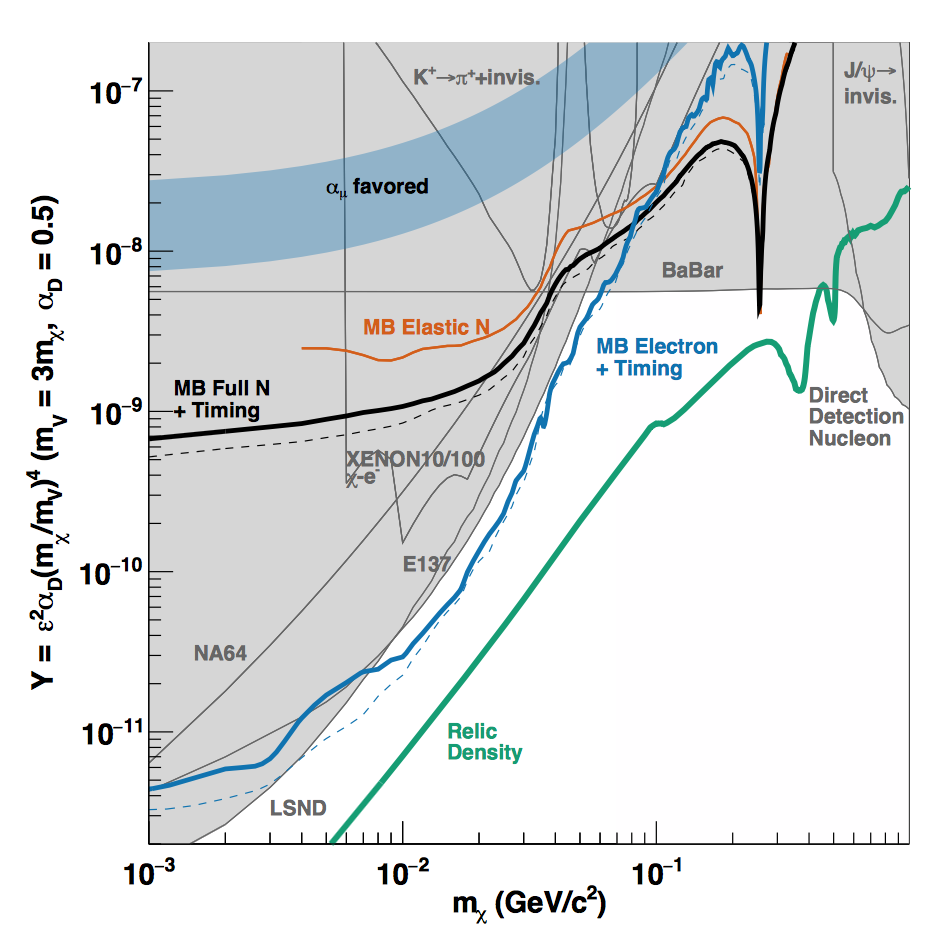}
\caption{\label{figure:MB-plot}
Results from the MiniBooNE-DM search for light dark matter from Ref.~\cite{Aguilar-Arevalo:2018wea}
}
\end{center}
\end{figure}

{\bf Light Dark Matter:} 
Accelerator based neutrino oscillation experiments have significant sensitivity to light dark matter~\cite{Batell:2009di,deNiverville:2011it,deNiverville:2012ij,Dharmapalan:2012xp,Batell:2014yra,Dobrescu:2014ita,Soper:2014ska,Kahn:2014sra,deNiverville:2015mwa,Coloma:2015pih,deNiverville:2016rqh,Izaguirre:2017bqb,Frugiuele:2017zvx,Jordan:2018gcd,deNiverville:2018dbu,Batell:2018fqo,Pospelov:2017kep,Ge:2017mcq}. The basic experimental principle is simple: light dark matter particles are produced in the proton-target collisions and subsequently travel to a detector downstream where they can scatter with electrons or nuclei, leaving a neutral current-like signature. This capability has recently been demonstrated in a dedicated search by MiniBooNE~\cite{Aguilar-Arevalo:2018wea,Aguilar-Arevalo:2017mqx}, which placed new limits on the well-motivated vector portal dark matter model~\cite{Pospelov:2007mp}, as shown in Figure~\ref{figure:MB-plot}.
There are a number of future research directions in the search for light dark matter at accelerator based neutrino experiments. Besides the  simplest vector portal model, variations of the dark sector models utilizing different mediators can be explored, see e.g.,~\cite{Batell:2014yra,Krnjaic:2015mbs,Kahn:2018cqs,Cherry:2014xra,Bertoni:2014mva,Batell:2017cmf}. 
Furthermore, there can be striking signatures beyond the basic neutral-current-like scattering signal, see e.g.,~\cite{Izaguirre:2017bqb,deGouvea:2018cfv,Kelly:2019wow}.
Looking beyond the MiniBooNE experiment, dark matter searches can be carried out at NO$\nu$A ~\cite{Dobrescu:2014ita,deNiverville:2018dbu}, MicroBooNE~\cite{Coloma:2015pih}, JSNS$^2$~\cite{Jordan:2018gcd}, DUNE~\cite{DeRomeri:2019kic,Brown:2018rcz}, and COHERENT~\cite{deNiverville:2015mwa,Dutta:2019nbn}, among others.
%
%

{\bf Dark sector mediators:} It is also possible to directly probe the mediator particle of the dark sector in accelerator- and reactor-based neutrino experiments. There are a number of well motivated possibilities for the mediator particle and interactions, including dark photons (vector portal), dark scalars (Higgs portal), heavy neutral leptons (neutrino portal), pseudo-scalars (axion portal), new gauge bosons ($B-L$, $L_\mu-L_\tau$). One strategy applicable to neutrino detectors is to search for the visible decays of the mediator particle to SM final  states~\cite{Batell:2009di,Essig:2010gu,Pospelov:2017kep,Tsai:2019mtm}. This complements other probes such as rare meson decays, electron beam fixed target experiments, B-factories, and astrophysical sources such as e.g., SN1987A (for a broad overview, see Ref.~\cite{Alexander:2016aln}). In a related direction, new strategies and dedicated experiments to search for long lived particles at CERN can also probe dark sector mediators~\cite{Beacham:2019nyx,Alimena:2019zri}. 

\subsubsection{Boosted Dark Matter}
\label{sec:bdm}
%
{\bf Elastic BDM:} 
As discussed in Sec.~\ref{sec:exec-landscape}, a component of dark matter could be boosted in the present universe. 
In the 
boosted dark matter (BDM) model with two components~\cite{Belanger:2011ww,Agashe:2014yua}, for example, the lighter component $\chi_1$ produced by pair-annihilation of the heavier component $\chi_0$ acquires a large Lorentz boost factor given by the mass ratio of the two components. 
Semi-annihilating dark matter (charged under e.g., $Z_3$ symmetry) can be decently boosted if the other annihilation product is light enough~\cite{DEramo:2010keq}.   
If such BDM has a sizable coupling to SM 
particles, it may leave relativistic scattering signatures at terrestrial detectors. 
The simplest possibility is its elastic scattering-off either electron or nucleon, i.e., $\chi_1 + e^-/N \to \chi_1 + e^-/N$. 
Since the expected flux of BDM from annihilation is generally small (suppressed by $n_{\rm DM}^2$)~\cite{Agashe:2014yua}, large-volume detectors are desirable.
Prospective signal sensitivities are investigated in Super-K/Hyper-K~\cite{ Agashe:2014yua, Kong:2014mia, Huang:2013xfa, Necib:2016aez, Alhazmi:2016qcs}, DUNE~\cite{Necib:2016aez, Alhazmi:2016qcs}, IceCube/PINGU~\cite{Agashe:2014yua, Kong:2014mia}, and LUX~\cite{Cherry:2015oca}. 
Surface-based detectors such as ProtoDUNE and SBN detectors may have sensitivity for $m_{\chi_0} \lesssim \mathcal O ({\rm GeV})$, restricting to the upward-going signals~\cite{Kim:2018veo}.

The signature of elastic BDM bears background from atmospheric-neutrino-induced events. Such neutrino background can be reduced by using directional information as BDM typically come from DM concentrated region such as the galactic center and the Sun. Relativistic nature of incident $\chi_1$ renders typical recoil target moving in the forward direction so that the associated signal source can be traced back in track-based detectors.
Therefore, an angle cut allows to focus on signal-rich regions hence improve signal sensitivities. 
Example studies include Sun-originating BDM~\cite{Berger:2014sqa, Kong:2014mia, Alhazmi:2016qcs} via the annihilation of solar-captured heavier dark matter and dwarf-galaxy-originating BDM~\cite{Necib:2016aez}. Further background rejection for elastic scattering may be achieved by dedicated analysis on event kinematics and using the fact that unlike neutrinos, BDM events have no correlated charged current events \cite{Agashe:2014yua}.

Variations and extensions of 
BDM models may lead to specific signals that are easier to distinguish from neutrino background. Some examples are described as follows.

\begin{figure}
\centerline{
\includegraphics[width=0.48\linewidth]{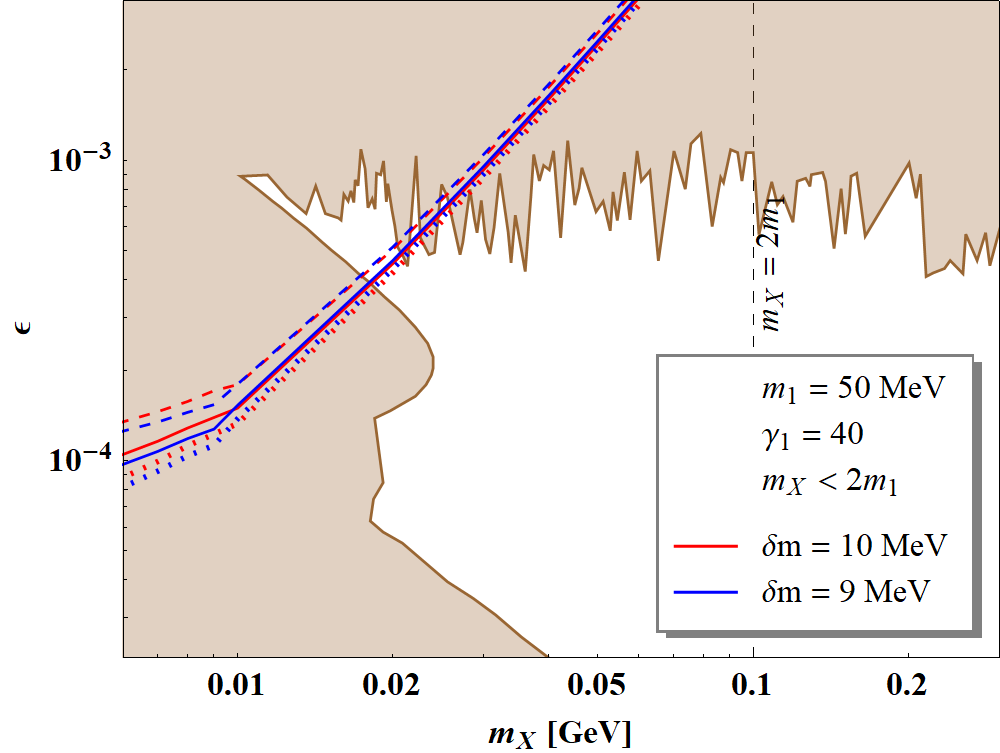}
\includegraphics[width=0.48\linewidth]{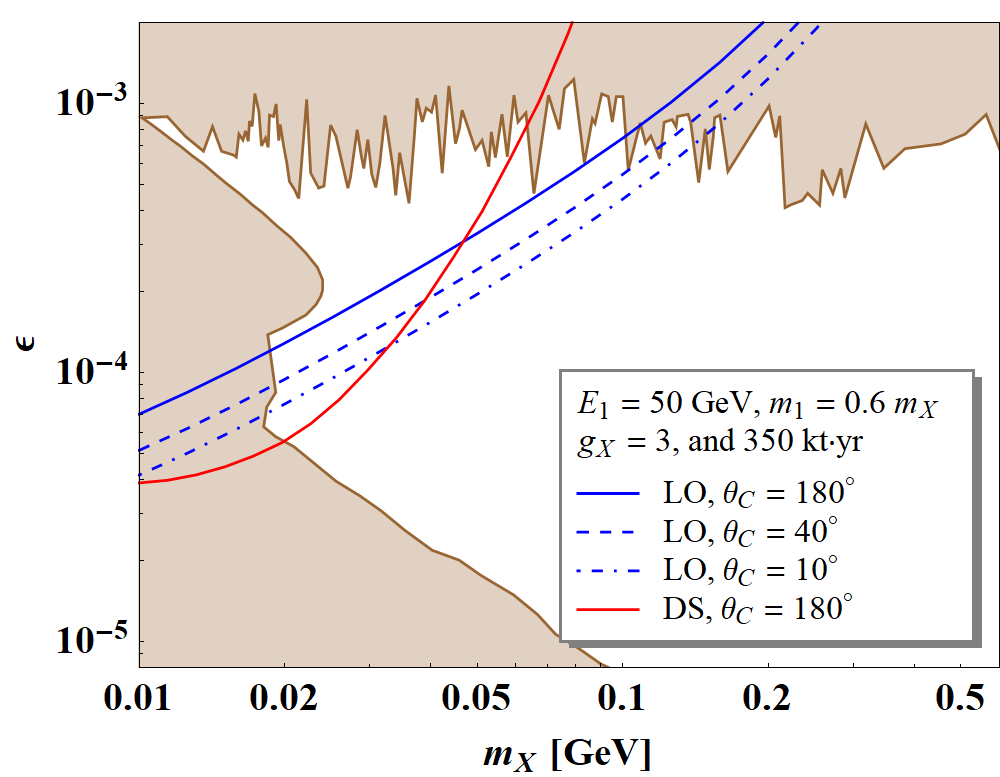}
}
\caption{Experimental sensitivities of searching for iBDM  at ProtoDUNE (left) and dark-strahlung process at DUNE (right). See Refs.~\cite{Chatterjee:2018mej,Kim:2019had} for the details.}
\label{fig:iBDM}
\end{figure}

{\bf Inelastic BDM (iBDM):}
If the dark sector of interest contains additional, presumably unstable state, say $\chi_2$ heavier than $\chi_1$ and boosted $\chi_1$ is allowed to up-scatter to $\chi_2$.
By construction, $\chi_2$ disintegrates back into $\chi_1$ potentially along with additional visible particles, e.g., an $e^-e^+$ pair. 
Due to the existence of extra features, the associated signal may suffer from substantially less background contamination, and as a consequence, great signal sensitivities would be achieved.
Associated signal searches have been proposed in various experiments such as Super-K/Hyper-K~\cite{Kim:2016zjx}, DUNE~\cite{Kim:2016zjx}, Xenon1T/LZ/DEAP3600~\cite{Giudice:2017zke}, and ProtoDUNE~\cite{Chatterjee:2018mej}. 
The senstivities of ProtoDUNE in searching for electron-recoiling iBDM signal are shown in the left panel of Fig.~\ref{fig:iBDM}.

{\bf BDM with dark-strahlung:}
In the scenarios in which dark photon mediates the interactions between the dark-sector and SM 
particles, the best background rejection may be achieved by searching for the signal processes
involving additional dark photon radiation (namely dark-strahlung) since similar higher-order processes are (nearly) unavailable in the neutrino scenario~\cite{Kim:2019had}.
Indeed, the dark-strahlung process gets more efficient in increasing energy of BDM and cosmogenic BDM can be energetic with a decent signal flux. 
Thus, this channel can then provide complementarity to the leading-order process, as shown in the right panel of Fig.~\ref{fig:iBDM}.

{\bf Expectations for BDM searches in existing neutrino experiments:} 
As the existing BDM searches are conducted at completely different sizes of detectors by SK~\cite{Kachulis:2017nci} and COSINE-100~\cite{Ha:2018obm}, other large-volume next-generation detectors are potentially capable of testing BDM/iBDM models. 
Available are electron and nucleon scattering channels. Depending on the energy transfer, the target nucleon may break apart, hence BDM-induced deep inelastic scattering processes come into play. 
BDM-induced DIS processes are included in the latest version of GENIE~\cite{Berger:2018urf} 
, and some useful guiding principles in searches for BDM/iBDM are elaborated in Ref.~\cite{evep}. 
The anomalous ultrahigh energy events in ANtartic Impulsive Transient Antenna (ANITA) is also explained by an iBDM-like model~\cite{Heurtier:2019rkz}. 
\subsubsection{Millicharged Particles}
%
The search for millicharged particles (MCP), or fractionally charged particles, is closely related to fundamental mysteries in nature: electric charge quantization and the nature of dark matter.
Once a U(1) dark gauge symmetry exists unbroken, we can take a field basis that dark matter is charged under U(1)$_{\rm EM}$~\cite{Holdom:1985ag} while the SM electromagnetic interactions remain unchanged. 
Due to the properties of DM and experimental constraints, the EM charge of DM is smaller than $\lesssim 10^{-3}$ of the electron charge.  

The search for MCP is especially challenging for the mass regime above MeV, where current lab, astrophysical, and cosmological bounds run out \cite{Davidson:2000hf, Chang:2018rso}.
Traditionally, the probe of this mass range is done in electron facilities and collider experiments \cite{Davidson:2000hf,CMS:2012xi,Agnese:2014vxh}. 
A dedicated experiment, milliQan, is proposed at LHC 
\cite{Haas:2014dda}, and can reach unprecedented sensitivity.
The possibility of MCP search in the neutrino facilities has long been proposed \cite{Golowich1987,Babu:1993yh,Gninenko:2006fi}. In the past, a dedicated search  
was performed in Kamiokande-II~\cite{Mori:1990kw} water Cherenkov experiment. 
Large volume neutrino experiments can be sensitive to MCP dark matter, accelerated in astrophysical sources, with competitive limits placed by SK~\cite{Hu:2016xas}.

In Ref.~\cite{Magill:2018tbb}, searches for MCP in numerous past (LSND, MiniBooNE, and MiniBooNE DM run), current (MicroBooNE), and future (SBND, DUNE and SHiP) proton fixed target and neutrino experiment was performed.  A more involved background analysis based on double-hit signatures at ArgoNeuT is discussed in \cite{Acciarri:2018myr,Harnik:2019zee}. Recently, a dedicated experiment, FerMINI~\cite{Kelly:2018brz}, is proposed utilizing the neutrino facilities and a dedicated scintillator-based detector to search for MCP.


\section{Tasks and Timelines}
\label{sec:exec-tasks}

With the array of models and tools already available, robust analyses can and have been developed at current experiments.  
Nevertheless, given so many areas of opportunity for BSM searches at current and upcoming neutrino experiments, it is imperative that efforts in this direction be increased.  
As novel search strategies are conceived, new challenges will be made clear for the neutrino community.
These, in turn, will contribute to a better understanding of the capabilities of relatively new detector technologies, such as LArTPCs, opening further areas of opportunity for upcoming experiments.

A crucial requirement to the success of the narrative above is the development of better interfaces between theory and experiment, and this is the focus of the present section.
The development of tools to aid experimental searches in neutrino experiments has received reinvigorated effort on recent years, mostly due to the unprecedented precision of forthcoming oscillation experiments.
Better modeling of neutrino-nucleus interactions, typically made via neutrino event generators, are necessary to carry out a precision neutrino physics program, as standard model processes are usually the major backgrounds in new physics searches.
For instance, in short and long baseline neutrino oscillation experiments, unknown nuclear physics dynamics and non-perturbative effects lead to large uncertainties; while in coherent  neutrino-nucleus scattering measurements, quenching effects and nuclear form factors represent the main source of systematics.

Besides improving the modeling of standard physics effects, the integration of BSM physics scenarios into neutrino event generators and other automated tools is extremely important for the development of robust experimental search strategies.
In fact, neutrino Monte Carlo codes, such as \verb+GENIE+~\cite{Andreopoulos:2009rq,Andreopoulos:2015wxa}, \verb+NEUT+~\cite{Hayato:2009zz}, \verb+NuWRO+~\cite{Golan:2012wx}, and \verb+GiBUU+~\cite{Buss:2011mx} already incorporate some new physics scenarios.
Efforts in this direction have been made recently and have led to substantial progress as, for instance, $n\overline{n}$ oscillations and boosted dark matter scenarios in \verb+GENIE+, the implementation of fixed-target event generation in \verb+MadGraph+ via \verb+MadDump+, and the implementation of dark sector generators in dedicated beam Monte Carlo codes like \verb+g4numi+ and \verb+g4bnb+.  
In the coming year, further improvements to the BSM capabilities of neutrino event generators should be made, with the implementation of missing nuclear processes, as well as new BSM models such as the dark neutrino and inelastic boosted dark matter scenarios.  

The simulations of beam and atmospheric neutrino fluxes  should be made more flexible and more robust, allowing for new BSM models as they develop and including better modeling of heavy flavor processes.  
In the case of atmospheric neutrinos, for example,  flux uncertainties play a crucial role in  sterile neutrino analysis such as the one performed by the  IceCube collaboration, while the neutrino events themselves are the main backgrounds for boosted dark matter searches.
In neutrino beam experiments, several BSM searches are limited by flux modeling uncertainties, such as nonstandard interactions and sterile neutrino searches exploiting both near and far detectors.
Following the example of BSM searches at the LHC and other colliders, the development of fast simulation tools adequate to BSM searches would be extremely valuable to further advance the field of neutrino physics.

Given the timeline of the Snowmass process, namely late 2021 for having the report ready, some of the aforementioned efforts should be prioritized.
{\bf By 2020 or early 2021}, several BSM scenarios should be implemented in neutrino event generators.
These include gauge-mediated and scalar-mediated  boosted dark matter scenarios (the former is already implemented in \verb+GENIE+),  
inelastic boosted dark matter, and dark neutrino scenarios.
Moreover, to estimate the experimental sensitivity to these BSM scenarios in a robust way, a standard ``pipeline'' from BSM \verb+ROOT+~\cite{Brun:1997pa} event files to detector simulation frameworks (such as \verb+LArSoft+~\cite{Church:2013hea}) and more realistic beam-related BSM simulations (e.g. including effects of magnetic horns in \verb+MadDump+ via \verb+g4bnb+ and \verb+g4numi+) should be available in early {\bf 2021}.
Meeting such a timeline is important to provide robust sensitivity studies of future neutrino efforts to several new physics scenarios for the Snowmass process. 
These studies will be incorporated into a final document with the intent of delineating the full capabilities and clearly explaining the physics case of such future neutrino efforts to the broad physics community. Finally, models of neutrino-nucleus interactions, whose impacts will permeate the entire standard and beyond standard neutrino physics program, will be improved continuously  throughout the next decade, specially after the completion of the SBN program~\cite{Machado:2019oxb}.~\footnote{Other important areas in neutrino physics such as neutrinoless double beta decay will be covered in a different white paper. Considerations about improvements in detector technology can be found in Sec.~\ref{sec:exec-nextgen}.}



\section{Conclusions}
\label{sec:exec-conclusions}

Over the last twenty years, extraordinary discoveries in neutrino physics have led to the development of new, intense neutrino sources and new, very sensitive and very large neutrino detectors. 
The unique capabilities of these sources and detectors allow one to pursue a rich and exciting program that extends beyond the pursuit of the physics of neutrino oscillations.

Here, we provide a summary of current, near-future, and future neutrino sources and detectors, concentrating on the characteristics of the different detector technologies and the impact these might have in the pursuit of new phenomena. 
As far as new phenomena are concerned, we focus on (1) the possibility that neutrinos have new, unexpected properties, or participate in interactions beyond those dictated by the Standard Model, and (2) the possibility that dark matter produced either in nature or in the beam-dump-like set-ups that produce intense neutrino fluxes can be studied with large neutrino detectors. 

The search for new phenomena in neutrino experiments, broadly defined, is a very active and vibrant area of current particle physics phenomenology. In this document, we summarized a variety of existing research efforts along several different directions that may lead to new discoveries at the next-generation of neutrino experiments. 
We also discussed the computational and physics tools available for understanding the search of new phenomena in neutrino experiments and tried to identify outstanding practical issues that need to be addressed in the near-future if the community is to optimize the physics reach of these types of experiments.

We further expect this document will help facilitate future workshops on additional probes and topics, such as nucleon decay or neutrinoless double beta decay, providing both a general forum and reference for discussion of BSM Physics searches at future neutrino experiments within the context of the upcoming decadal study of the APS Division of Particles and Fields.


\section{Acknowledgements}
\label{sec:exec-acknowledgements}
The authors would like to thank the University of Texas at Arlington, the Division of Particles and Fields of the American Physical Society and the U.S. Department of Energy (award number DE-SC0019742) for providing supports for the Workshop on New Opportunities at the Next Generation Neutrino Experiments held on April 12 and 13, 2019 at the University of Texas at Arlington, which this white paper stems from.
The authors also would like to thank the following colleagues who have endorsed this paper while they are not authors: J. Asaadi (U. of Texas at Arlington), S.K.Agarwalla (IOPB), C. Giunti (INFN Torino), S. Goswami (PRL Ahmedabad), D. Grant (Michigan State University), P. Huber (Virginia Tech), S.B.Kim (Seoul National University), and S. Petkov (SISSA/INFN and IPMU).



\renewcommand{\refname}{References}

\printglossary

\bibliographystyle{utphys}

\bibliography{common/tdr-citedb}

\end{document}